\documentclass[pre,amsmath,amssymb,showpacs,twocolumn]{revtex4}

\usepackage{subfigure}
\usepackage{epsfig}
\usepackage{amsmath}
\usepackage{graphics}
\usepackage{graphicx}
\usepackage{times}
\usepackage{color}

\begin{document}
\title{ Nonlinear Driven Response of a Phase-Field Crystal in a Periodic Pinning Potential}

\author{C. V. Achim $^1$, J.A.P. Ramos$^{2,3}$, M. Karttunen $^4$, K.R. Elder$^5$, E. Granato$^{3,6}$,
T. Ala-Nissila$^{1,6}$ and S.C. Ying$^6$}
\address{$^1$Department of Applied Physics, Helsinki University of
Technology, P.O. Box 1100, FIN-02015 TKK, Espoo, Finland}
\address{$^2$ Departamento de Ci\^encias Exatas, Universidade
Estadual do Sudoeste da Bahia, 45000-000 Vit\'oria da Conquista,
BA,Brazil}
\address{$^3$Laborat\'orio Associado de Sensores e Materiais,
Instituto Nacional de Pesquisas Espaciais, S\~ao Jos\'e dos Campos,
SP Brazil}
\address{$^4$Department of Applied Mathematics, The University of
Western Ontario, London (ON), Canada N6A\,5B7}
\address{$^5$Department of Physics, Oakland University, Rochester,
Michigan, 48309-4487, USA}
\address{$^6$Department of Physics, Brown University, Providence,
R.I. 02912-1843, U.S.A.}

\date{\today}

\begin{abstract}
We study numerically the phase diagram and the response under a
driving force of the phase field crystal model for pinned lattice
systems introduced recently for both one and two dimensional
systems. The model describes the lattice system as a continuous
density field in the presence of a periodic  pinning potential,
allowing for both elastic and plastic
deformations of the lattice.  We first  present  results for phase
diagrams of the model in the absence of a driving force. The
nonlinear response to a driving force on an initially pinned
commensurate phase is then studied via overdamped dynamic
equations of motion for different values of mismatch and pinning
strengths. For large pinning strength the driven depinning
transitions are continuous, and the sliding velocity varies with
the force from the threshold with power-law exponents in agreement
with analytical predictions. Transverse depinning transitions in
the moving state are also found in two dimensions. Surprisingly,
for sufficiently weak pinning potential we find a discontinuous
depinning transition with hysteresis even in one dimension under
overdamped dynamics. We also characterize structural changes of
the system in some detail close to the depinning transition.

\end{abstract}


\maketitle{}


\section{Introduction}

There exist many systems in nature with two or more competing length
scales, which often leads to the appearance of spatially modulated structures.
Such systems may exhibit both commensurate (C) and incommensurate (I)
phases \cite{bib:shick81,bib:Bak82be} characterized by differences in the
spatial ordering of the system. Important examples include spin density
waves \cite {bib:Ruvalds85,bib:Faw88}, charge density waves
\cite{bib:Fleming85ky}, vortex lattices in superconducting films with
pinning centers \cite{bib:Martin97} and weakly adsorbed monolayers
\cite{bib:Thomy80pl,bib:Pandit82tu} on a substrate. The emerging
structures are characterized by an order parameter (\emph{e.g.} charge, spin
or particle density) that is modulated in space with a given wave vector $q$.
In particular, for 2D adsorbate systems, there is competition between the
commensurate state which is favored by a strong periodic pinning potential
and the cost of the elastic energy depending on the mismatch between the
intrinsic lattice constant $a$ of the overlayer, and the period $b$ of the
pinning potential.

While the static properties of C and I structures have been
extensively characterized \cite{bib:shick81,bib:Bak82be} much less
is known about their dynamics. A particularly interesting case
arises when an initially  pinned phase is subjected to an external
driving force $f$. The resulting nonlinear response is relevant for
a variety of different physical systems which are accessible
experimentally. A driven atomic monolayer in a periodic pinning
potential is an interesting realization of such nonlinear behavior
\cite{bib:Perssonbook}, which is directly relevant for experiments
on sliding friction between two surfaces with a lubricant
\cite{bib:Israelechvili} and between adsorbed layers and an
oscillating substrate \cite{bib:Krim,bib:Mistura}. Other systems of
great interest are driven charge density waves
\cite{bib:Fisher83,bib:Fisher85,bib:Gruner81} in which
commensurability and impurity pinning often compete \cite{bib:tomic97},
and superconductor vortex arrays in which different commensurability
and pinning behaviors have been experimentally
observed~\cite{bib:welp05,bib:wu05,bib:villegas}, including
periodic and asymmetric potentials.

For a sufficiently
large pinning potential, the phase may remain pinned for small
forces if there are no thermal fluctuations present. This means that
at zero temperature there is a finite critical force $f_c$ above
which the system starts moving. For many systems, it is found that
just above the threshold $f_c$, the drift velocity $v_d$ shows a
power-law dependence  with respect to the force $f$
\begin{equation}
v_d\propto (f-f_c)^{\zeta}.
\end{equation}
When this behavior is regarded as a dynamical critical phenomenon,
the power-law exponent $\zeta$ of the corresponding driven depinning
transition can be argued to result from the scaling behavior of the
system near the threshold \cite{bib:Fisher83} with corresponding
divergent time and length scales and universal behavior. In general,
however, the observed value of $\zeta$ may depend on the system and
its dimensionality. For a pure elastic medium with quenched
randomness there appears to be a universal value which depends on
the dimensionality of the system, provided inertial effects are
negligible \cite{bib:Middleton91,bib:Middleton92}. For the case of
an initially commensurate phase in a periodic pinning potential
without disorder, a power-law exponent $\zeta = 1/2$ is expected
independent of the dimension  as the threshold behavior can be
understood from the point of view of single particle behavior
\cite{bib:Fisher83,bib:Fisher85,bib:Gruner81,bib:Sethna93}. In the
limit of a large force $(f-f_c)/f_c \gg 1$ the system is moving and
the corresponding relationship between the driving force and the
sliding velocity defines the sliding friction coefficient
\begin{equation}\label{eq:slfr}
\eta_s = f/v_d.
\end{equation}

The simple Frenkel-Kontorova (FK) model
\cite{bib:Kontorova39wq,bib:ChLu95} extended to two dimensions and
other similar elastic models have been used to study driven
depinning transitions and the sliding friction of adsorbed
monolayers \cite{bib:Perssonbook,bib:Granato99}. Although these
models take into account topological defects in the form of domain
walls they leave out plastic deformations of the layer due to other
defects such as dislocations. These defects are particularly
important when the CI transition occurs between two different
crystal structures, or in presence of thermal fluctuations or
quenched disorder, and should be taken into account for a more
realistic description of the system. Such defects can be
automatically included in a full microscopic model involving
interacting atoms in the presence of a substrate potential using
more realistic interaction potentials. However, the full
complexities of the microscopic model severely limit the system
sizes that can be studied numerically, even when simple
Lennard-Jones potentials are used to describe the interactions
\cite{bib:Persson93,bib:Granato00}.

When the driving depinning transition is discontinuous, hysteresis
effects can occur which result in two different critical forces,
$f_{c}^{in}> f_{c}^{de}$, corresponding to the threshold values for
increasing the force from zero and decreasing the force from a large
value, respectively.
A fundamental issue in modeling such systems is the origin of the
hysteresis. It is well known that hysteresis can occur in
underdamped systems, where inertia effects are present. However,
molecular dynamics simulations of a 2D model of an adsorbed layer
with Lennard-Jones interacting potential for increasing values of
the damping coefficient (microscopic friction) and analytical
arguments suggested that hysteresis should remain in the overdamped
limit \cite{bib:Perssonbook,bib:Perssonchem}. On the other hand,
results for the pure elastic FK model shows that although the
hysteresis behavior is similar to Lennard-Jones model for weak
damping, it disappears in the overdamped limit \cite{bib:Granato99}.
The different behaviors could be due to the absence of some defects
generated during the depinning transitions, which are allowed in the
Lennard-Jones model but not in the elastic FK model. In fact,
hysteresis can be argued to arise from topological defects in the
lattice such as dislocations in two-dimensional systems
\cite{bib:Fisher85} even in the absence of inertial effects. One
thus expects that overdamped dynamics should be able to described
the hysteresis behavior in two dimensions provided the model
incorporates both elastic and plastic deformations. For charge
density waves, where the pinning potential is disordered, a field
theoretical model has been introduced which allows for dislocations
as well as thermal fluctuations \cite{bib:Kart99} through amplitude
and phase fluctuations, and shows both elastic and hysteretic
behavior in agreement with experiments \cite{bib:rring}. In absence
of disorder, however, the possible hysteresis behavior in such
models has not been investigated.

Recently a phase field crystal (PFC) model was introduced
\cite{bib:Elder02kz, bib:Elder04rq,bib:Elder07} that allows for
both elastic and plastic
deformations in the solid phase. In this formulation a free energy functional
is introduced which depends on the field $\psi(\vec{r},t)$ that corresponds to
the particle number density averaged over microscopic times scales. The free
energy is minimized when $\psi$ is spatially periodic (\emph{i.e.},
crystalline) in the solid phase and constant in the liquid phase. By
incorporating phenomena on atomic length scales the model naturally
includes elastic and plastic deformations, multiple crystal orientations and
anisotropic structures in a manner similar to other microscopic approaches
such as molecular dynamics. However, the PFC model describes the density
on a diffusive and not the real microscopic times scales. It is therefore
computationally much more efficient.

In our previous works \cite{bib:Achim06au,bib:Achim08ap} we
demonstrated how the influence of an external periodic pinning
potential can be incorporated in the PFC model. Such a model
provides a continuum description of pinned lattice systems. The
pinning potential is chosen such that it allows the occurrence of
both C and I phases as ground states of different symmetries in
the model. In Ref. \cite{bib:Achim06au}  part of the phase diagram
as function of pinning strength and lattice mismatch between the
pinning potential and the PFC was mapped out. Numerical
minimization was used to find the minimum free-energy
configurations and provide details on the topological defects in
the boundary region. In particular, we found that the transition
from the I to the C phase remains discontinuous for all values of
the mismatch studied in Ref. \cite{bib:Achim06au}. We also
performed a detailed Voronoi analysis of the defects throughout
the transition region. In Ref. \cite{bib:Achim08ap} the
equilibration method was improved and the range of mismatches
extended to include both positive and negative mismatches.

In the present work we focus on the case where the  PFC under an
external periodic potential without disorder introduced in Ref.
\cite{bib:Achim06au} is driven by an external force in the absence
of thermal fluctuations. To this end, we first present improved
detailed phase diagrams of the model both in 1D and 2D. The main
focus of the present work is on the influence of an external driving
force on the pinned C phase, which we study for different values of
mismatch and pinning strengths for 1D and 2D systems. As expected,
due to the competition between the pinning potential and the driving
force there is a depinning transition at $f_c$ for a finite driving
force $f$. We demonstrate that within a certain range of parameters
the depinning transitions are continuous, and find that both in 1D
and 2D the corresponding power-law exponent is $\zeta =0.5$ in
agreement with the expected value
\cite{bib:Fisher83,bib:Fisher85,bib:Gruner81,bib:Sethna93}. We also
characterize structural changes of the system close to the depinning
transition. For large pinning strength transverse depinning
transitions in the moving state are also found. Surprisingly, for
sufficiently weak pinning potential we find a discontinuous
depinning transition with hysteresis even in one dimension although
overdamped dynamical equations are used.

\section{The Phase Field Crystal under  a periodic potential and  a Driving Force }

For the phase field crystal in the presence of pinning potential
\cite{bib:Elder02kz, bib:Elder04rq,bib:Elder07,bib:Achim06au}, the
free energy functional can be written in dimensionless form as %
\begin{equation}\label{eq:dimF}
F =\int
d\vec{x}\left[\frac{\psi}{2}\left(r+\left(1+\nabla^2\right)^2\right)\psi+\frac{\psi^4}{4}+V(\vec
x)\psi(\vec x)\right],
\end{equation}
where $r$ is a temperature dependent quantity and $V(\vec x)$
is an external potential which represents the effect of the
substrate.  This model can be
derived directly from the classical density functional theory
\cite{bib:Elder07} of freezing by expanding around the
properties of a liquid in coexistence with a solid phase.  More
specifically it can be shown \cite{bib:Huang08} that
$r$ is proportional to the difference between the isothermal
compressibility of the liquid and the elastic energy of
crystalline phase.  Furthermore the length scales in
this model have been scaled by the nearest neighbour
distance in the coexisting liquid state.

In the absence of the pinning potential the equilibrium minimum
energy configuration of the system depends on the parameter
$r$ and the average density
$\bar\psi=\frac{1}{V_d}\int d\vec x \psi(\vec x)$
\cite{bib:Elder04rq}, where $V_d$ is the system volume in $d$
dimensions. In 2D, the solid phase corresponds to a triangular
lattice. The length scale chosen here corresponds to $k_0 ^{-1}$,
where $k_0 =2\pi/(a_t\sqrt{3}/2)=1$, with $a_t$ as the lattice
constant of the intrinsic triangular lattice. When the external
pinning potential is present, the competition between the length
scales associated with the intrinsic ordering and the pinning
potential can lead to complicated phases depending on the
parameters chosen in the energy functional \cite{bib:Achim06au,bib:Achim08ap}.
We choose an external pinning potential of a simple periodic form
$V=V_0\cos(k_sx)$ in 1D, and $V=V_0[\cos(k_sx)+\cos(k_sy)]$ in 2D.
The wave vector $k_s$ is related to the periodicity of the pinning
potential $a_s$, such that $k_s=2\pi/a_s$.

We define the relative mismatch
$\delta_\mathrm{m}$ between the external potential and the PFC as
\begin{equation}
\delta_\mathrm{m}=(1-k_s)
\end{equation}
The response of the system to a driving force can be obtained by
including a convective derivative, $\vec{f}\cdot\vec{\nabla}\psi$,
to the  original PFC model \cite{bib:Elder04rq}. Thus, the
dynamical equation of motion for the phase field is given by
\begin{eqnarray}
\frac{\partial\psi}{\partial\tau}&= & \nabla^2 \frac{\delta
F}{\delta \psi}+ \vec{f}\cdot{\vec{\nabla}}\psi \cr &=&
\nabla^2\left(\left(r+\left(1+\nabla^2\right)^2\right)\psi
+\psi^3+V\right)  \cr &
& + \vec{f}\cdot\vec{\nabla}\psi.
\label{eq:tdgleq}
\end{eqnarray}
For the forces considered in this work (typically
$\vec{f}=f\hat{x}$), the convective term does not change the average
value of the density field.

In contrast to the usual classical microscopic characterization of
particle positions and velocities, the measurement of an average
drift velocity in response to an external driving force $\vec{f}$
requires some discussion. In the PFC model the maxima of the density
field that define the lattice structure cannot always be interpreted
as individual particles, since vacancies may be present in the
system. The conservation law in the model concerns the local density
field, not the number of maxima in the field. This becomes evident
in the driven PFC model, where the motion of the density field close
to depinning may be more akin to flow in a continuous medium than
the motion of discrete particle-like objects. Thus, defining the
drift velocity in terms of the density field maxima is
computationally more demanding to implement.  We have found that
measuring the drift velocity $v_d$  from the rate of change of the
gradient of the density field gives consistent results,
in absence of thermal fluctuations. We have used the following
definition

\begin{equation}
v_d\equiv\left<{\left<\left|{\partial \psi }/{\partial
t}\right|\right>_{\vec x}}\,\big/{\left<\left|{\partial
\psi}/{\partial x}\right|\right>_{\vec x}}\right>_t,
\label{velgrad}
\end{equation}
where the subscripts $\vec x$ and $t$ in the brackets denote
averaging over space and time, respectively.  Alternative
definitions were also considered, but this particular form proved
the most statistically accurate.

Although the definition of the drift velocity $v_d$ according to
Eq. (\ref{velgrad}) can be used to determine the velocity
response along the direction of the driving force, it is not
particularly useful in the study  of the response in the
transverse direction since it is not a vector quantity. In order
to study the transverse response it is more convenient to
determine the average velocity directly from the positions of the
local peaks in $\psi(\vec x)$. This requires locating such peaks
as a function of time during the numerical simulation. We have
developed a computational method which determines the location and
velocity of each individual peak in presence of the external force
and thermal fluctuations \cite{bib:vpeak}. The method to locate
the peaks is based on a particle location algorithm used in
digital image processing \cite{bib:Crocker}. The drift velocity
for the lattice of density peaks is obtained from the peak
velocities $\vec v_i$ as
\begin{equation}
\vec v_P = \left<\frac{1}{N_P}\sum_{i=1}^{N_P} \vec v_i(t)\right>_t,
\label{velpeak}
\end{equation}
where $N_P$ is the number of peaks. We find that the definitions of
the velocity from Eqs. (\ref{velgrad}) and (\ref{velpeak}) give
consistent results for the longitudinal depinning in absence of
thermal fluctuations.  However, in presence of thermal fluctuations
only the definition from the peak velocities (\ref{velpeak}) is able
to separate the contribution to the drift velocity  due to the
driving force from thermal noise contributions.

\section{Results}

In this section, we present  results obtained for the static and
dynamic properties of the PFC model. Numerically, we study the
system properties by integrating Eq. (\ref{eq:tdgleq}) using a simple
Euler algorithm and the time derivatives are approximated by a
forward finite difference with the time step $dt=0.005$ (the time
scale corresponds to the diffusion time over the length scale $k_0
^{-1}$). For the 1D case, the density field was discretized on a
uniform  grid with $dx=\pi/4$, while for the 2D case we used a
square uniform grid with $dx=dy=\pi/4$. The laplacians are
evaluated in 1D using a central difference, while for the 2D the
'spherical laplacian' is used \cite{bib:Elder04rq,bib:Patra05}. In
both the equilibrium and driven situations, fully periodic
boundary conditions have been used. Note that for the conserved
time-dependent Ginzburg-Landau (TDGL) equation (\ref{eq:tdgleq}),
the addition of the driving force of the form $f(\partial
\psi/\partial x)$ preserves the local conservation of $\bar\psi$
under periodic boundary conditions. For the study of the static
properties $\vec{f}$ is set to zero. For a given value of the
mismatch, the pinning strength $V_0$ is increased from zero to a
maximum in steps of $dV_0$ and then decreased back to zero. Each
time the pinning strength is changed we allow the system to
equilibrate. The final state corresponds to a configuration that
minimized the energy functional.

For the study of the influence of a driving force we choose the
mismatch and pinning strength such that the system is initially in
a commensurate state. The driving force $\vec{f}$ is then
increased from zero to a maximum value $|\vec{f}| > f_c$ and then
decreased back to zero. In the case where the depinning transition
at $f_c$ is continuous, we determine the corresponding depinning
exponent $\zeta$ in the limit $ |\vec{f}| \rightarrow f_c$. Unless
specified the force is applied in the $x$ direction, i.e.,
$\vec{f} = f \hat{x}$.

\subsection{Phase Diagram and Nonlinear Response  for 1D System}

\subsubsection{Equilibrium properties}
\begin{figure}[!h]
\begin{center}
\includegraphics[width=60mm,clip=true,angle=270]{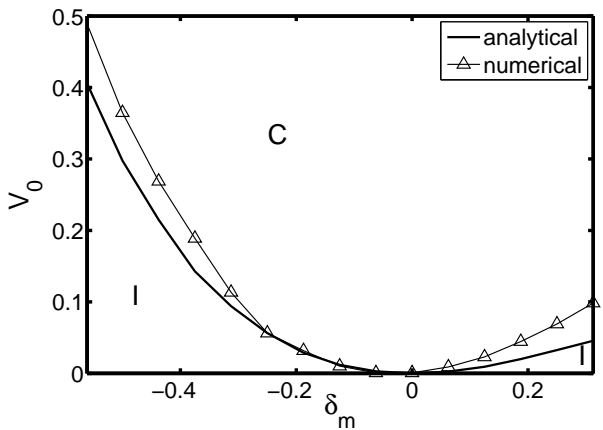}
\caption{\label{fig:OneDpd}
Phase boundary between commensurate (C) and incommensurate (I)
phases for the 1D pinned PFC model, calculated numerically
(continuous line with triangles) and analytically (continuous
line). }
\end{center}
\end{figure}

The static properties of the pinned PFC in 1D ($f=0$) are presented
as a phase diagram in the $V_0 - \delta_{\rm m}$ plane shown in Fig.
\ref{fig:OneDpd}. For comparison, we include in Fig.
\ref{fig:OneDpd} the phase boundaries obtained analytically and
numerically from from Eq. \ref{eq:tdgleq}. The analytical phase
boundary was obtained by minimizing the free energy $F[\psi(x)]$,
expanding the density field as
\begin{equation}
\psi(x)=A_1\cos(x)+A_2\cos(k_sx)+A_3\cos((2-k_s)x),
\end{equation}
where the last term accounts for the distortion of the
lattice.

\begin{figure}
\begin{center}
\subfigure[]
{\includegraphics[height=55mm,clip=true,angle=0]{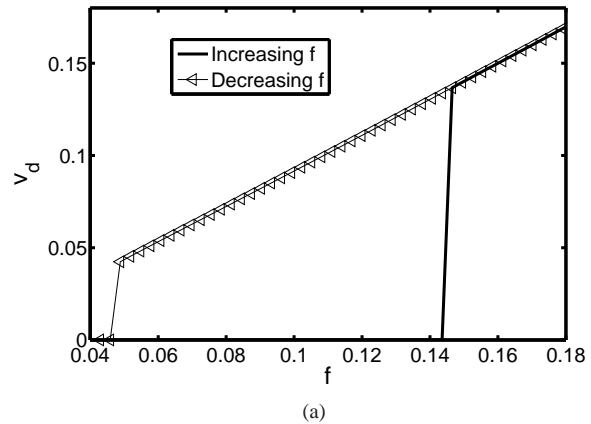}\label{subfig:220.1100v1}}
\subfigure[]
{\includegraphics[width=55mm,clip=true,angle=270]{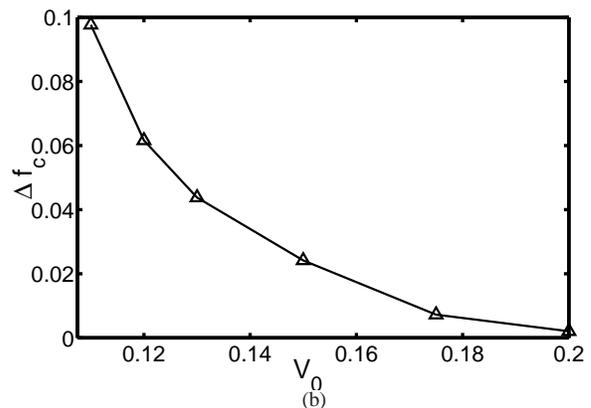}\label{subfig:deltafc-22}}
\end{center}
\caption{\label{fig:220.1100} (a) Discontinuous
depinning of the commensurate phase for relatively low pinning strength
($\delta_{\textrm{m}}=0.3125$, $V_0=0.11$). (b) Dependence of
$\Delta f_c$ on $V_0$ for $\delta_{\textrm{m}}=0.3125$.}
\end{figure}

Next we investigate the influence of an external force on the 1D
commensurate phase. For this purpose, the parameters are chosen
such that $r =-1/4$, and $\bar\psi=0$. Depending on the values of
the mismatch and pinning strength different behavior is found when
the driving force is added. Several values of mismatch between
$0.3125$ and $-0.50$ were investigated. For
$\delta_{\textrm{m}}\gtrsim -0.3$ two types of depinning are
present, discontinuous and continuous. For values of the pinning strength close to the I-C phase
transition, the depinning transition is discontinuous. The
dependence of the drift velocity as a function of the driving
force exhibits a hysteresis (see Fig. \ref{subfig:220.1100v1}).
The gap $\Delta f_c=f_c^{in}-f_c^{de}$ decreases when the pinning
strength increases (Fig. \ref{subfig:deltafc-22}), which indicates
that the transition becomes continuous for large enough $V_0$.

\begin{figure}[!]
\begin{center}
{\includegraphics[width=55mm,clip=true,angle=270]{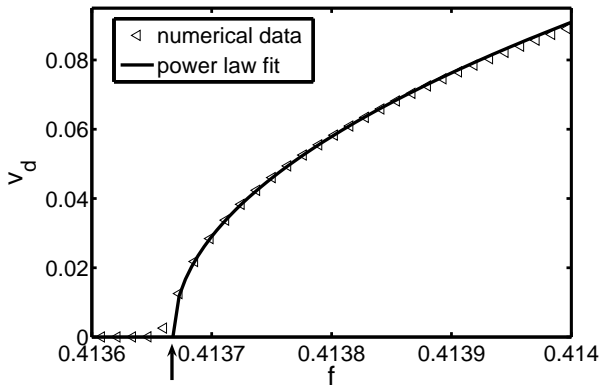}\label{subfig:220.2500v1}}
\end{center}
\caption{\label{fig:22cont}  Continuous depinning transition for relatively high pinning strength
($\delta_{\textrm{m}}=0.3125$ and $V_0=0.250$).
The arrow marks the value of the critical depinning force.
The triangles represent the numerical data while the continuous line is a fit to $(f-f_c)^{\zeta}$.
The best fit is obtained for the exponent $\zeta=0.50\pm 0.03$. }
\end{figure}

Note that for $\delta_{\textrm{m}}\lesssim -0.3$ the depinning
transition of a C pinned phase is continuous for all values of the
pinning strength, while for $\delta_{\textrm{m}}\gtrsim -0.3$ only
for values of pinning strength above a certain threshold. In Fig.
\ref{fig:22cont}, we show the behavior of a continuous depinning
transition for $\delta_{\textrm{m}}=0.3125$, $V_0=0.11$. The
dependence of the drift velocity on the force follows a power law
$v_d \propto (f-f_c)^\zeta$, as can be seen in Fig.
\ref{fig:22cont}. The exponent $\zeta$ does not depend of the
pinning strength and it is equal to $0.50\pm 0.03$ in all cases
studied here. This result can be understood as follows. When the
pinning potential is periodic and large in magnitude the neighboring
phases are weakly coupled and the system should behave as a single
particle in a periodic potential
\cite{bib:Fisher83,bib:Fisher85,bib:Gruner81,bib:Sethna93}. This
effective single particle behavior which is expected to describe the
threshold behavior for a commensurate phase in absence of defects,
is  independent of the dimension, leading to a depinning exponent
\cite{bib:Fisher83,bib:Fisher85,bib:Gruner81,bib:Sethna93}
$\zeta=1/2$. We also find that the critical force increases with the
pinning strength and for $\delta_{\textrm{m}}\gtrsim -0.3$ has
linear dependence on the pinning strength, while for
$\delta_{\textrm{m}}\lesssim -0.3$ its dependence on pinning becomes
sub-linear. Finally, we also note that for large driving forces $f
\gg f_c$ the system is totally depinned and the dependence of the
drift velocity on the force follows  Eq. (\ref{eq:slfr}) with a
linear dependence.

\subsection{Phase Diagram for 2D System}

\begin{figure}[!h]
\begin{center}
\subfigure[]{\includegraphics[width=29mm,clip=true]{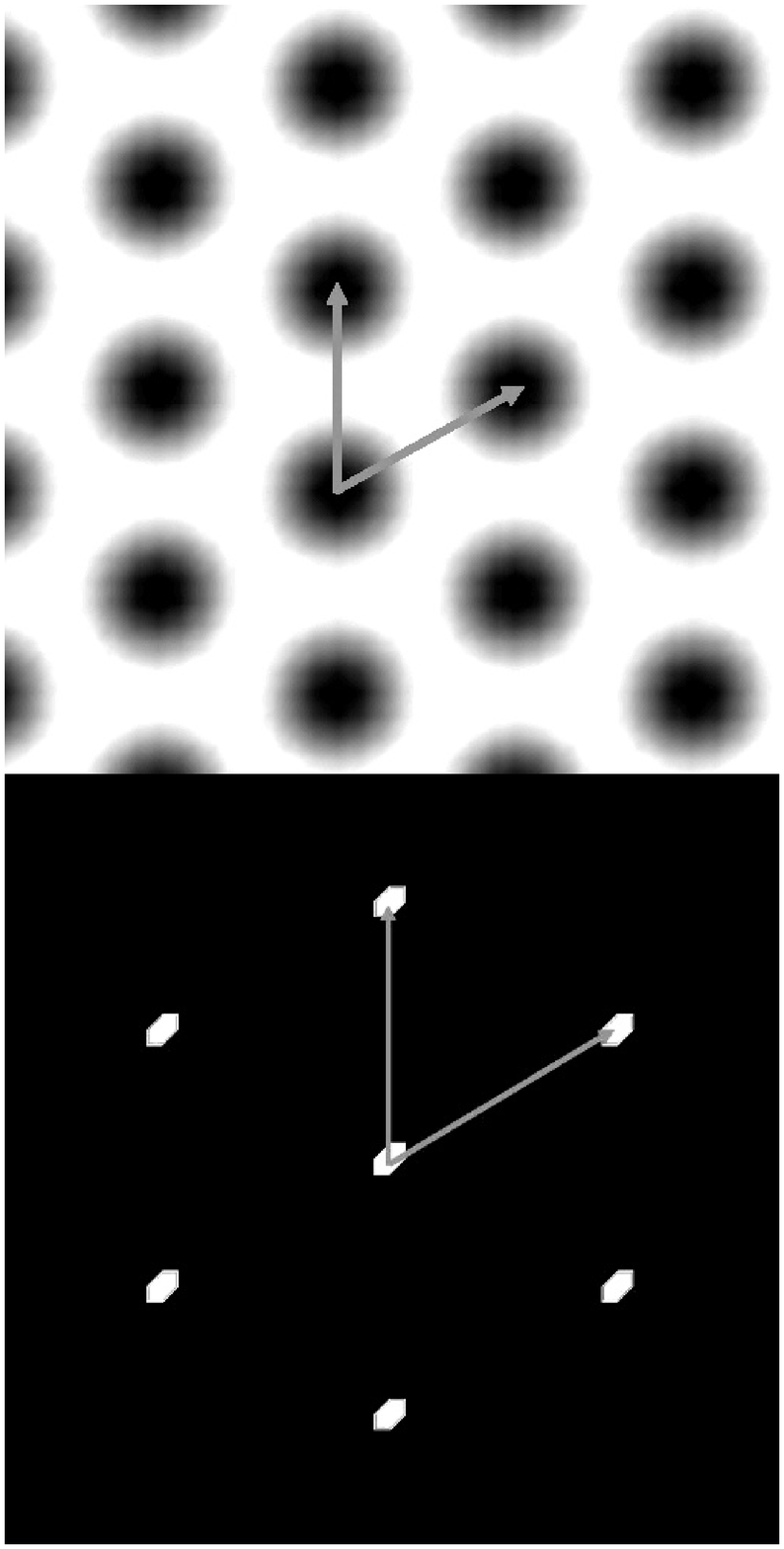}\label{hex}}
\subfigure[]{\includegraphics[width=29mm,clip=true]{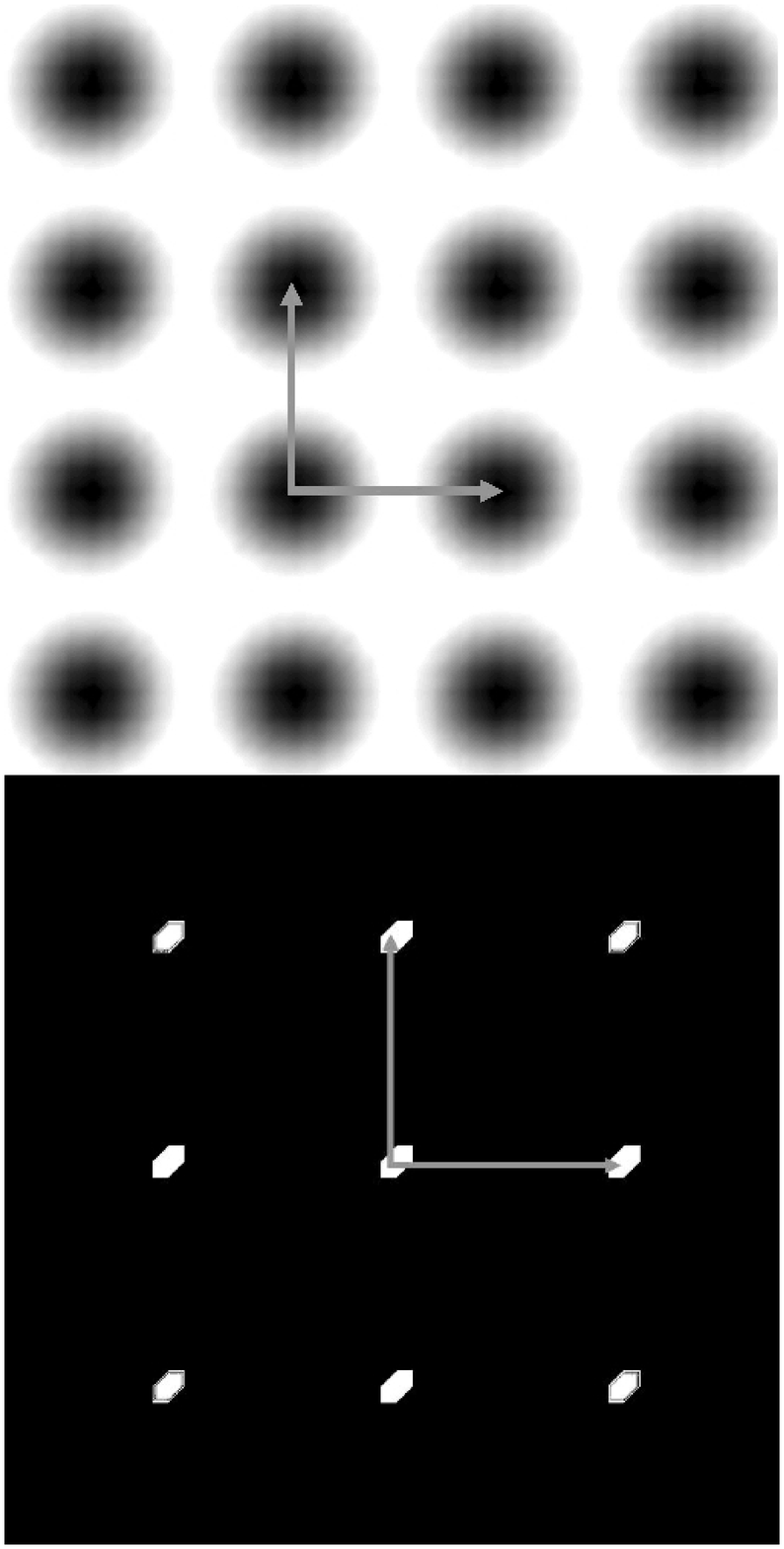}\label{square}}
\subfigure[]{\includegraphics[width=29mm,clip=true]{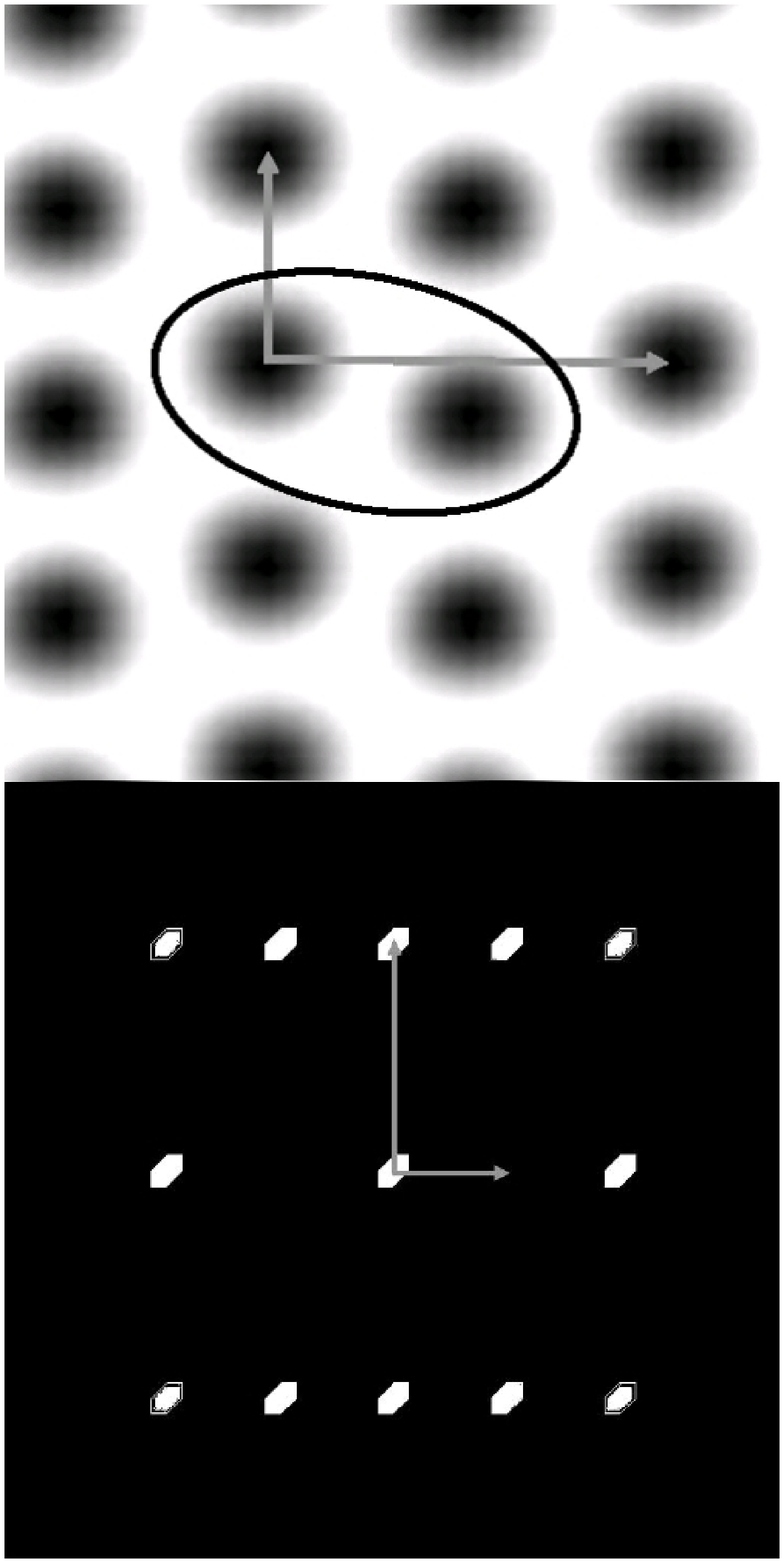}\label{square1}}
\subfigure[]{\includegraphics[width=29mm,clip=true]{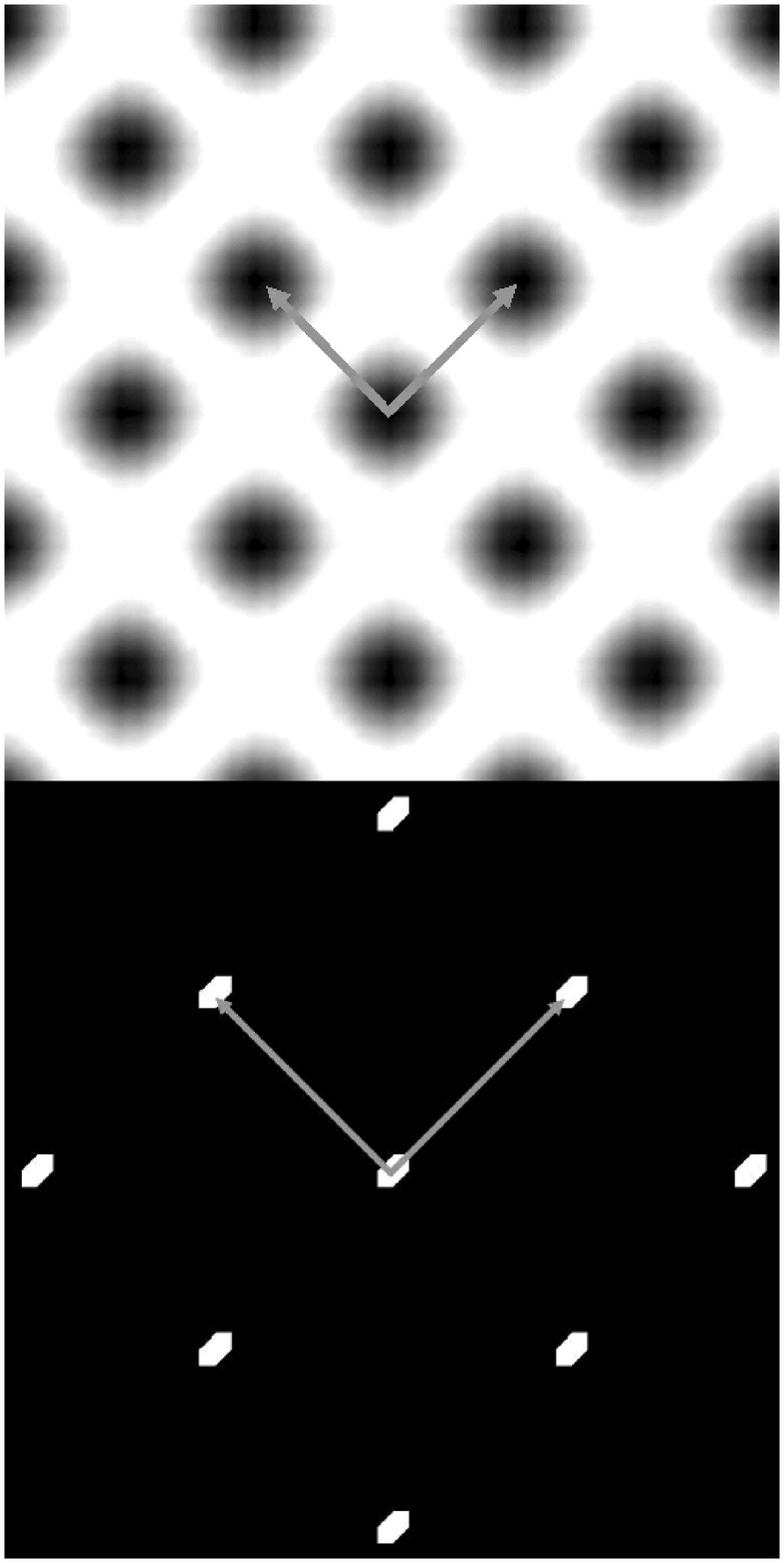}\label{square2}}
\subfigure[]{\includegraphics[width=29.4mm,clip=true]{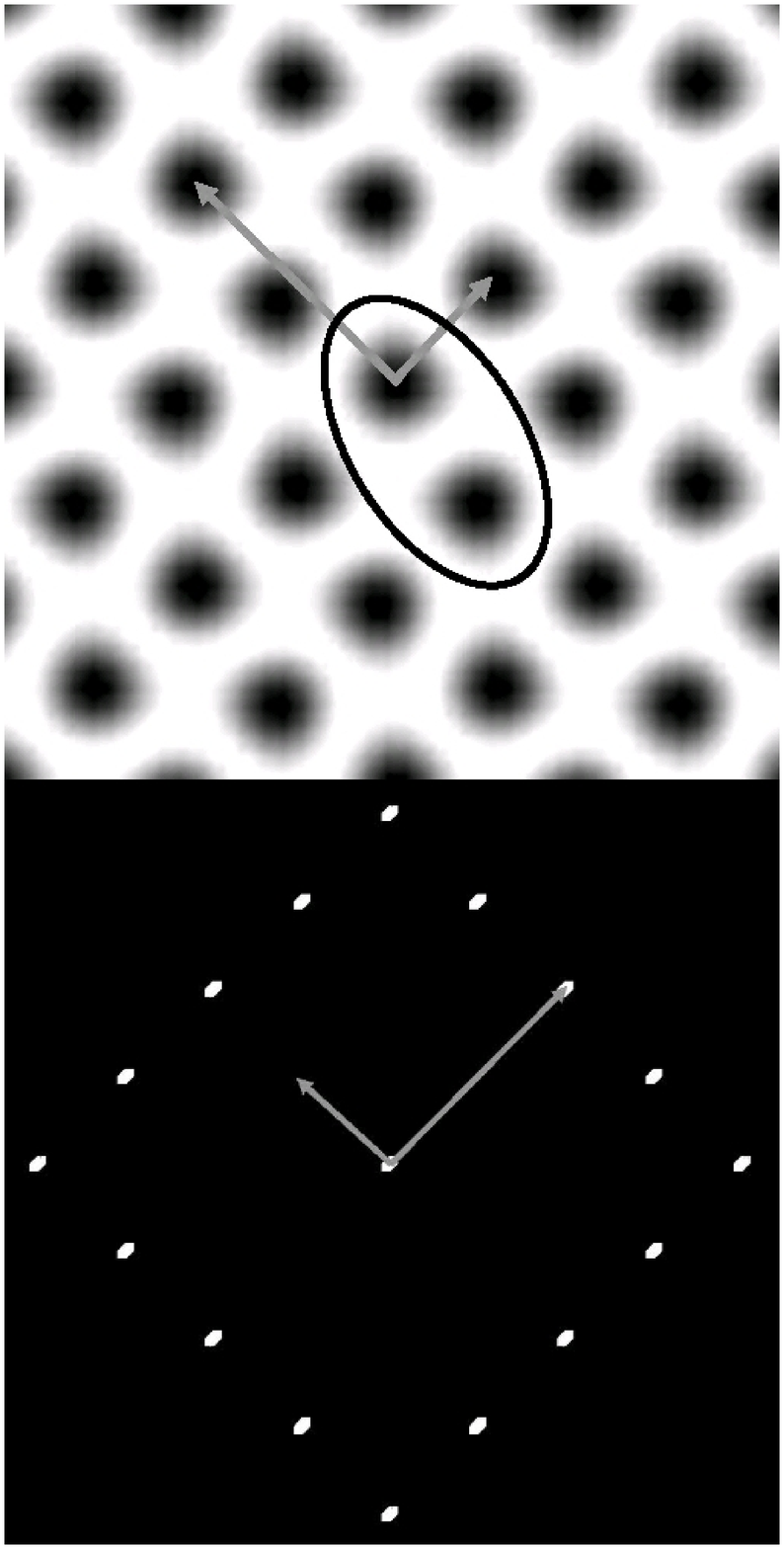}\label{square3}}
\end{center}
\caption{\label{phases} The phases that minimize the free-energy, according to Ref. \cite{bib:Achim08ap}: (a)
hexagonal, (b) square $(1\times1)$, (c) square $(2\times1)$,  (d)
square $\rm{c}(2\times2)$, and (e) square $2\sqrt{2}\times\sqrt{2}$.
The upper panels represent the density plotted in a gray colormap
and the corresponding lattice vectors, while the lower panels show the
structure factors and the relevant reciprocal lattice vectors. The
black contours in Figs. \ref{square1} and \ref{square3} show the
bases which generate the $(2\times1)$ and
$(2\sqrt{2}\times\sqrt{2})$ lattices.}
\end{figure}

Next we present a summary of the static properties of the 2D PFC
model in the presence of the external pinning potential
$V(x,y)=V_0(\cos(k_sx)+\cos(k_sy))$. The parameters chosen are
$r=-1/4$ and $\bar\psi=-1/4$. Analytically, we consider the
density to be a sum of hexagonal and square modes,
%
\begin{eqnarray}\label{eq:2danaexp}
\psi(x,y) & = & A_t (\cos(x\sqrt{3}/2) \cos(y/2)-\frac12\cos(y)) \\ \nonumber
& + & A_{s1}(\cos(k_sx)+\cos(k_sy)) \\ \nonumber
& + & A_{s2}\cos(k_sx)\cos(k_sy)\\ \nonumber
& + & A_c\cos(\frac{k_s}{2}x)\cos(\frac{k_s}{2}x)+\bar\psi.
\end{eqnarray}
For small values of the pinning strength
the system is in a hexagonal I phase for all mismatches (Fig. \ref{hex}). When the pinning strength is large enough, the system
will be in one of the commensurate phases, of which the
$(1\times1)$ phase is an exact match with the pinning potential
here (Fig. \ref{square}).

The other ordered phases are higher commensurate phases, which
exist only when one of the  reciprocal lattice vectors for the commensurate phase is close
to the wave vectors of the square pinning potential \cite{bib:Achim08ap}. One of these phases is the $\rm{c}(2\times 2)$ phase (Fig. \ref{square2}) in
which every second site of the lattice of the pinning potential corresponds to the
maximum in the phase field \cite{bib:shick81}.
This state is favored for mismatch values close to $1-\sqrt{2}$. Another higher commensurate phase is the
$(2\times1)$ (Fig. \ref{square1}) which is generated by a translation of the basis with
the reciprocal lattice vectors of the a $\rm{c}(2\times 1)$
lattice \cite{bib:shick81}. Finally, the
$(2\sqrt{2}\times\sqrt{2})$ phase (Fig. \ref{square3}) is similar to the $(2\times1)$
phase. The lattice is generated by a translation of the basis with
vectors which are rotated $45^{\circ}$ with respect to the pinning
potential and the magnitudes of the vectors are $2\sqrt{2}a_s$ and
$\sqrt{2}a_s$.
The phase is favored for mismatch values close to $-0.27$.

\begin{figure}[!h]
\begin{center}
\subfigure[]{\includegraphics[width=80mm,clip=true]{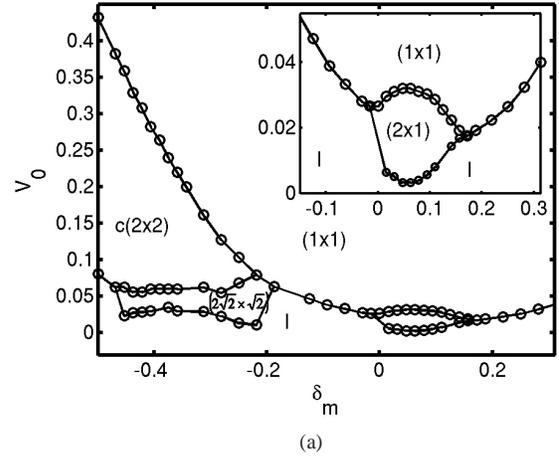}\label{subfig:pdnum}}
\subfigure[]{\includegraphics[width=80mm,clip=true]{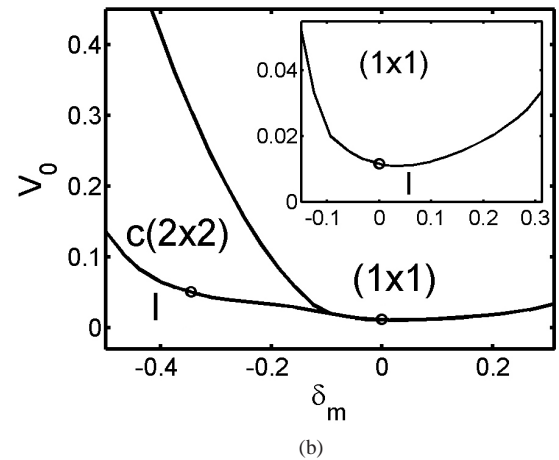}\label{subfig:pdana}}
\end{center}
\caption{\label{phases} The phase diagram in terms of the pinning
strength ($V_0$) and mismatch $\delta_{\rm m}$  calculated (a)
numerically, according to Ref. \cite{bib:Achim08ap} and (b)
analytically using approximation of the density given by Eq.
(\ref{eq:2danaexp}). The insets in (a) and (b) show the phase
diagram close to $\delta_{\textrm{m}}=0$. The circles in (b) mark the
values for which the approximation for the density given by Eq.
(\ref{eq:2danaexp}) breaks down.}
\end{figure}
The transitions between the different phases are found by
investigating the positions and the heights of the peaks in the
structure factor. The results of our extensive numerical
calculations are summarized in the phase diagram of Fig.
\ref{subfig:pdnum} which has been taken from  Ref.
\cite{bib:Achim08ap}.

\subsubsection{Nonlinear Response  of the $(1\times1)$ phase}

\begin{figure}
\begin{center}
\subfigure[]{\includegraphics[height=55mm,clip=true,angle=0]{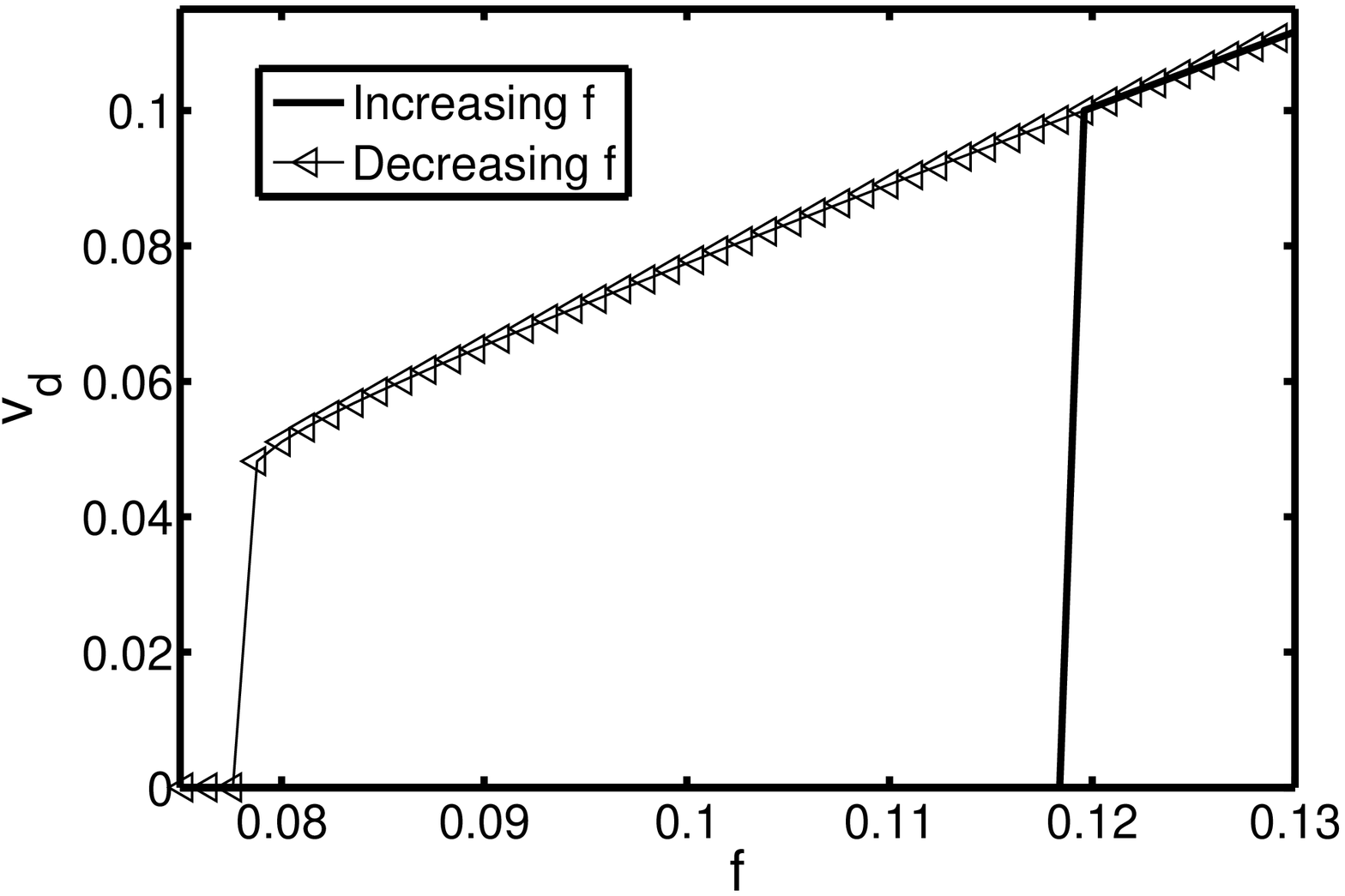}\label{subfig:128-28-0.0350v1}}
\subfigure[]{\includegraphics[width=55mm,clip=true,angle=270]{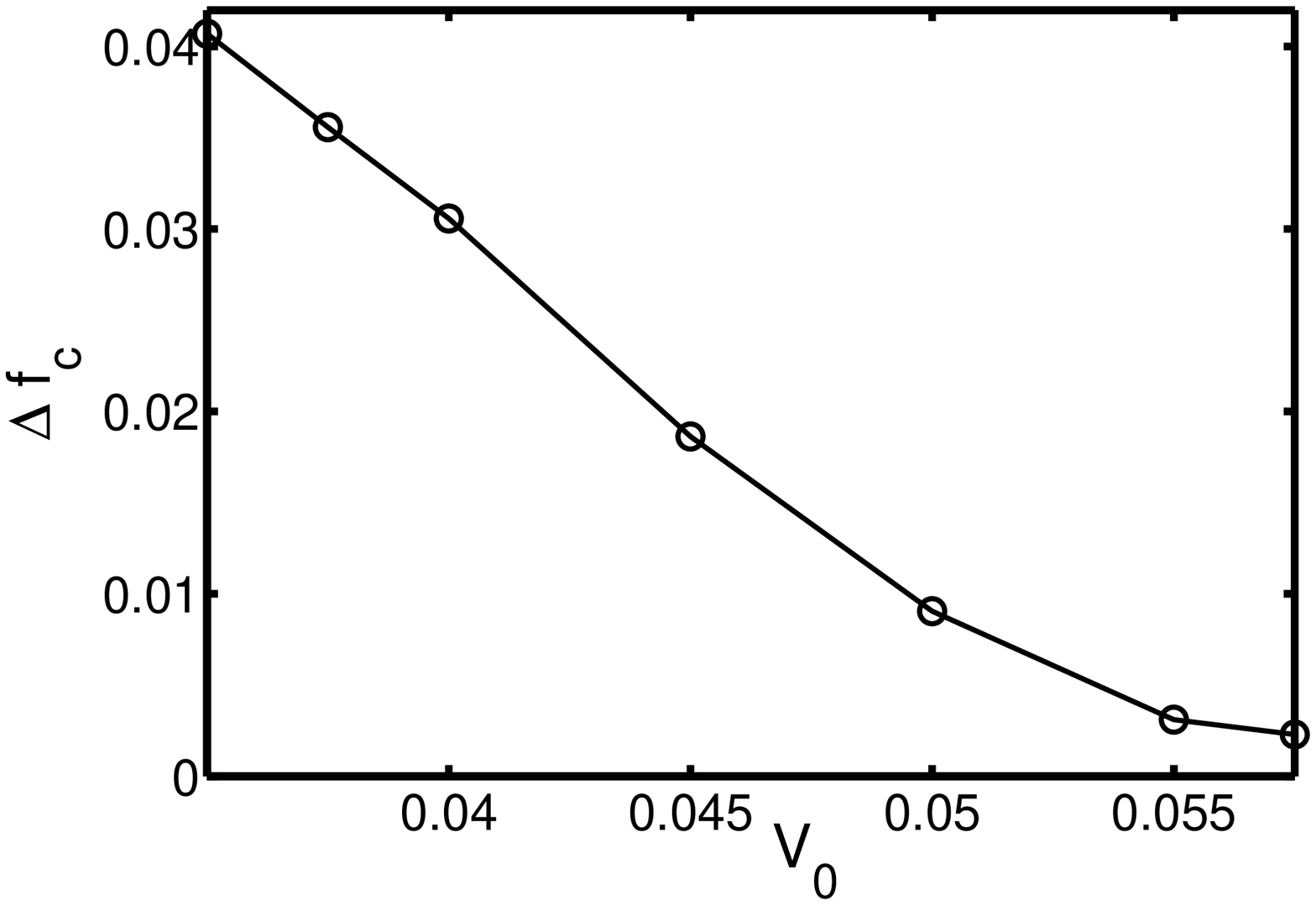}\label{fig:128-deltafc-28}}
\end{center}
\caption{\label{fig:128-28-0.0350} (a) The variation of the velocity
with respect to the external force for a discontinuous transition for the commensurate $(1\times 1)$ phase
($\delta_{\rm{m}}=0.125,V_0=0.0350$) and (b) the variation of
the gap $\Delta f_c$ vs $V_0$ for the same mismatch $\delta_{\textrm{m}}=0.125$. }
\end{figure}

\begin{figure}[!h]
\begin{center}
\subfigure[]{\includegraphics[width=55mm,clip=true,angle=270]{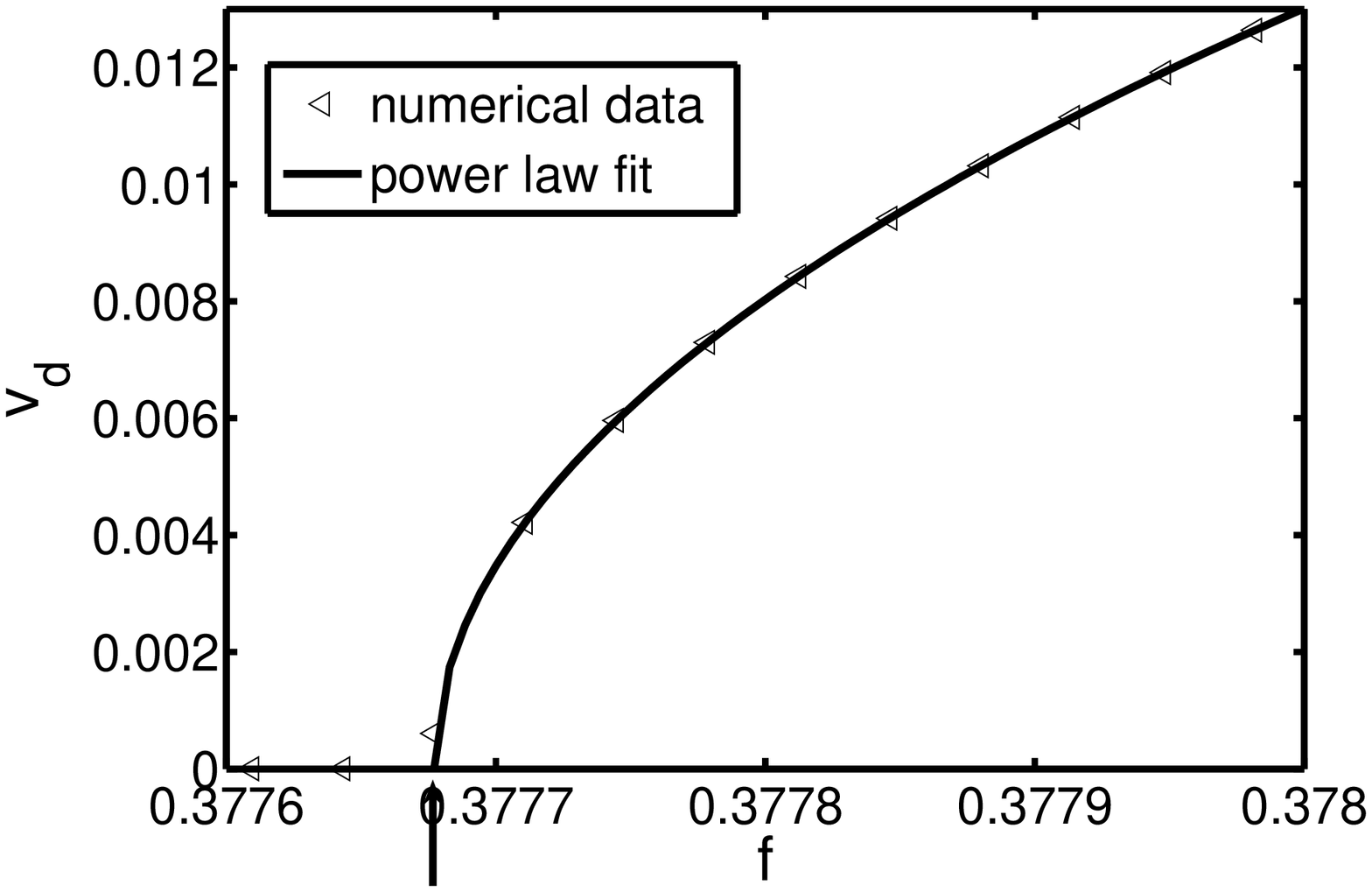}\label{subfig:128-28-0.0900v1}}
\end{center}
\caption{\label{fig:128-28-0.0900}  Dependence of the velocity
on the external force for a continuous transition for the $(1\times 1) $ phase ($\delta_{\rm{m}}=0.125,V_0=0.0900$).  The vertical arrow  marks the critical force $f_c$.
The triangles  represent the numerical data, while the continuous line is a power-law fit with $\zeta=0.50\pm 0.03$. }
\end{figure}

We now turn to the influence of an external driving force for the
pinned, commensurate $(1\times1)$ and $c(2\times2)$ phases.
Depending on the values of $\delta_{\rm{m}}$ and $V_0$, for both
phases we find both discontinuous and continuous depinning
transitions. For $\delta_{\rm{m}}\geq -0.2$ and $V_0 \geq 0.09$ both
continuous and discontinuous  depinning mechanisms were found for
the commensurate $(1\times1)$ phase. For smaller values of the
pinning strength close to the IC transition,  only discontinuous
depinning transitions were found. Similar to  the 1D case, we
identify two values of the critical forces for a discontinuous
depinning, namely $f_c^{in}$ for when the force is increased and
$f_c^{de}$ when the force is decreased back to zero. Figure
\ref{subfig:128-28-0.0350v1} shows the velocity dependence with
respect to the applied force for a discontinuous transition. We have
also tested the effect of thermal fluctuations here and find that
for temperatures low enough the hysteresis remains unchanged up to
some value which depends on the mismatch  and pinning strength. The
gap $\Delta f_c=f_c^{in}-f_c^{de}$ for a given $\delta_{\rm{m}}$
decreases as the pinning strength is increased (Fig.
\ref{fig:128-deltafc-28}). Finally, when the gap $\Delta f_c$
vanishes the depinning transition becomes continuous. In this regime
we find that the sliding velocity follows a power low $v_d\propto
(f-f_c)^{\zeta}$ (Fig. \ref{fig:128-28-0.0900}), consistent with
$\zeta=0.5$ in all cases studied here as in the 1D case. We note
that for $\delta_{\rm{m}}\leq -0.2$ only continuous depinning
transitions were found for all values of the pinning strength.

\begin{figure}[!h]
\begin{center}
\subfigure[]{\includegraphics[width=35mm,clip=true,angle=0]{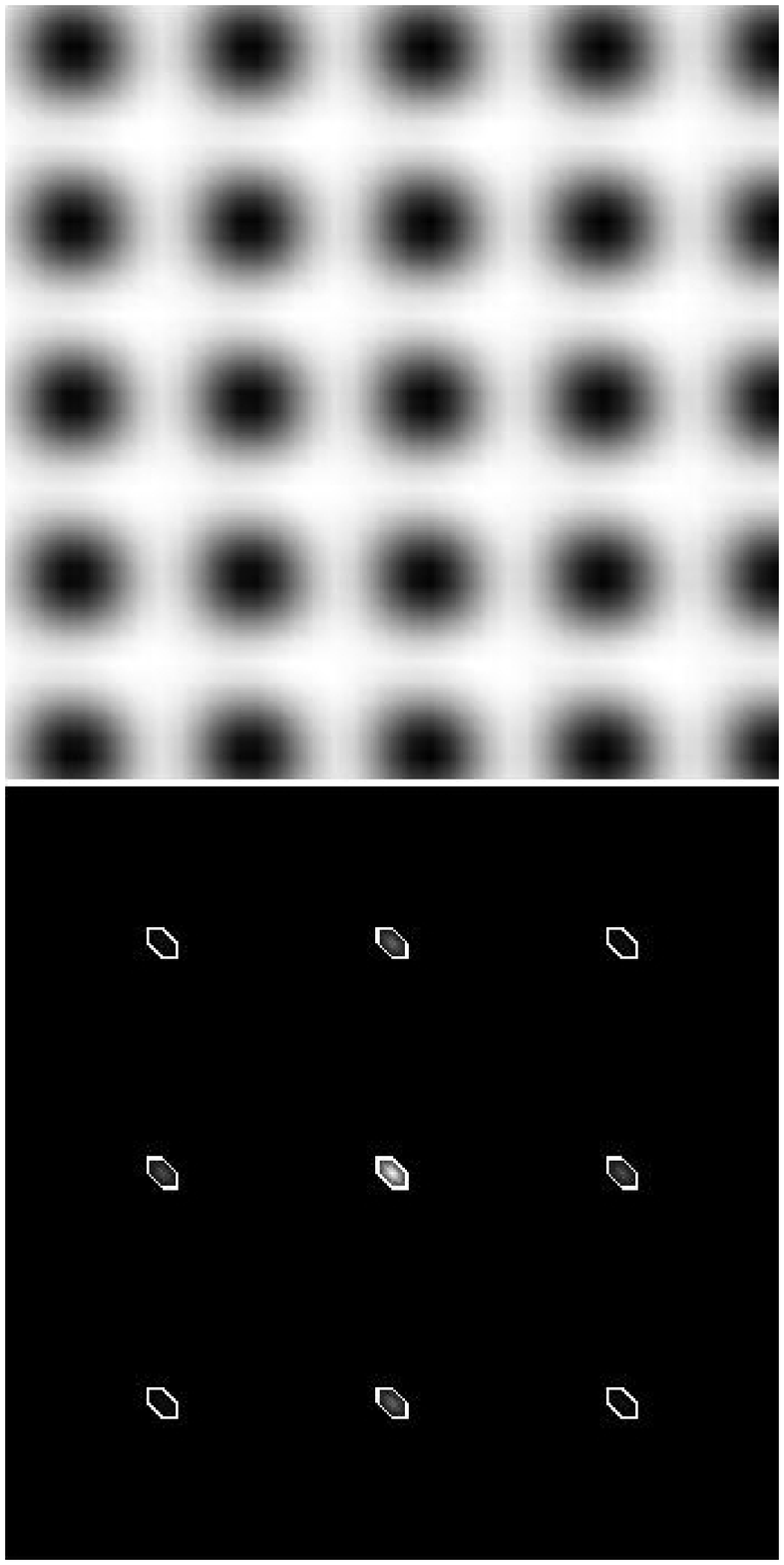}\label{subfig:vo0.0350f0.07040}}
\subfigure[]{\includegraphics[width=35mm,clip=true,angle=0]{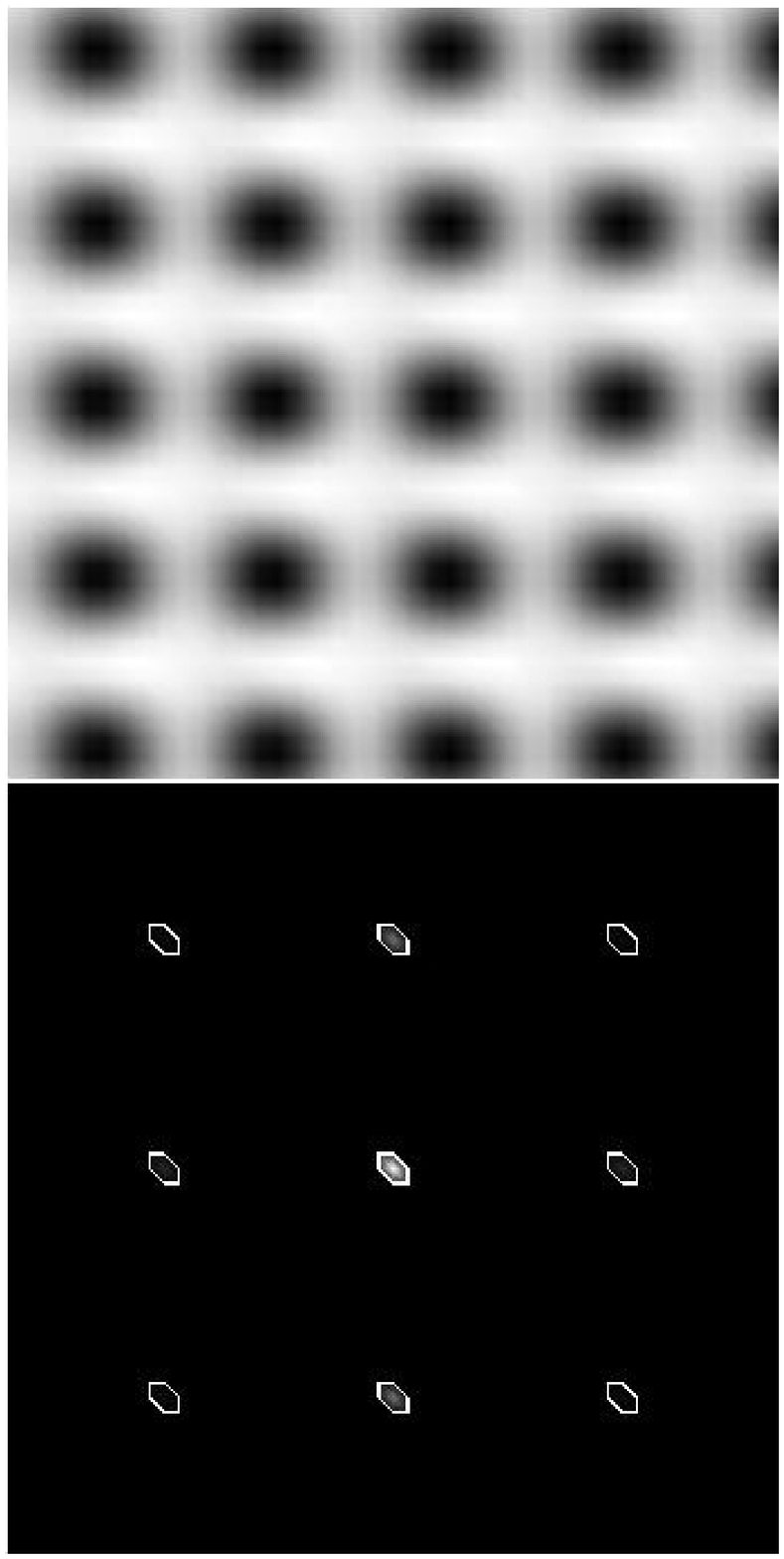}\label{subfig:vo0.0350f0.11840up}}
\subfigure[]{\includegraphics[width=35mm,clip=true,angle=0]{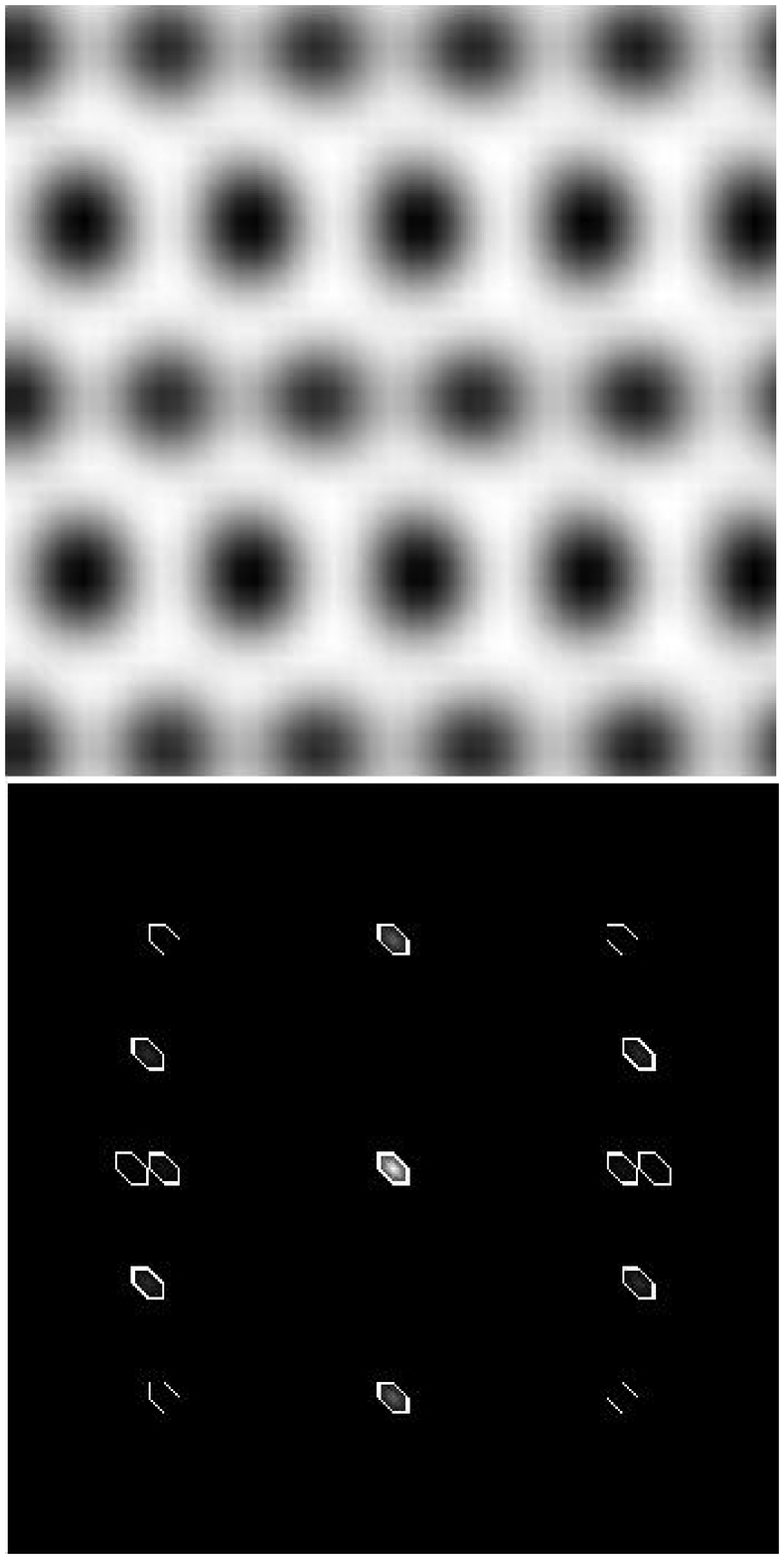}\label{subfig:vo0.0350f0.11000do}}
\subfigure[]{\includegraphics[width=35mm,clip=true,angle=0]{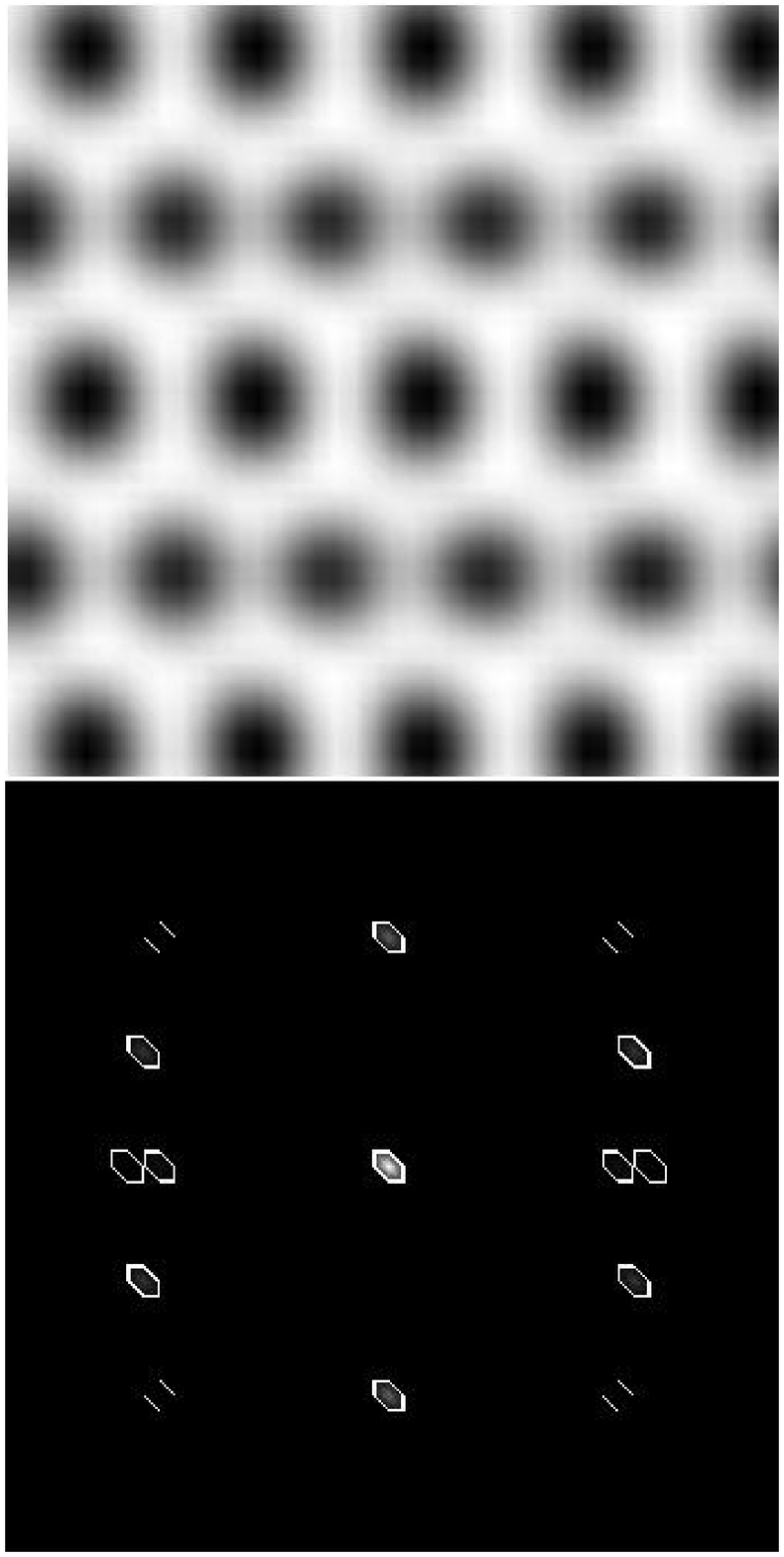}\label{subfig:vo0.0350f0.13040}}
\end{center}
\caption{Change of the lattice structure (upper panels) and the corresponding structure factor (lower panels) with the applied force for $\delta_{\rm{m}}=0.125,V_0=0.0350$, where depinning is discontinuous. Image (b)
corresponds to $f=0.11$ with a non-moving initial configuration, while for (c) the applied force is the same but the initial configuration is a moving one. The case (a) $f=0.07$, (d) $f=0.13$ are outside of the hysteresis
region and same result is obtained with moving or non-moving initial configuration. }\label{fig:vo0.0350f}
\end{figure}

\begin{figure}[!h]
\begin{center}
\subfigure[]{\includegraphics[width=35mm,clip=true,angle=0]{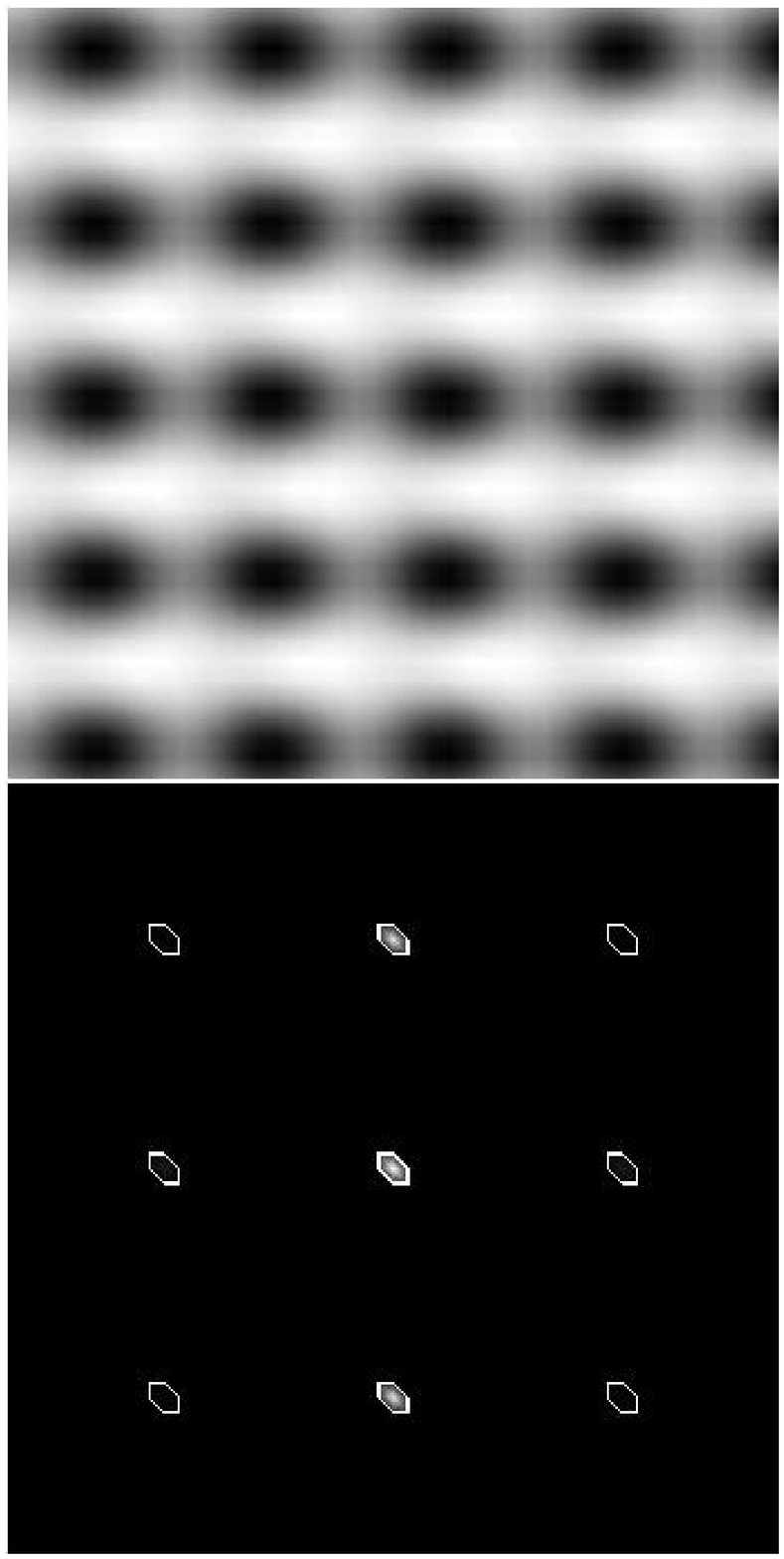}\label{subfig:vo0.0900f0.3776432000}}
\subfigure[]{\includegraphics[width=35mm,clip=true,angle=0]{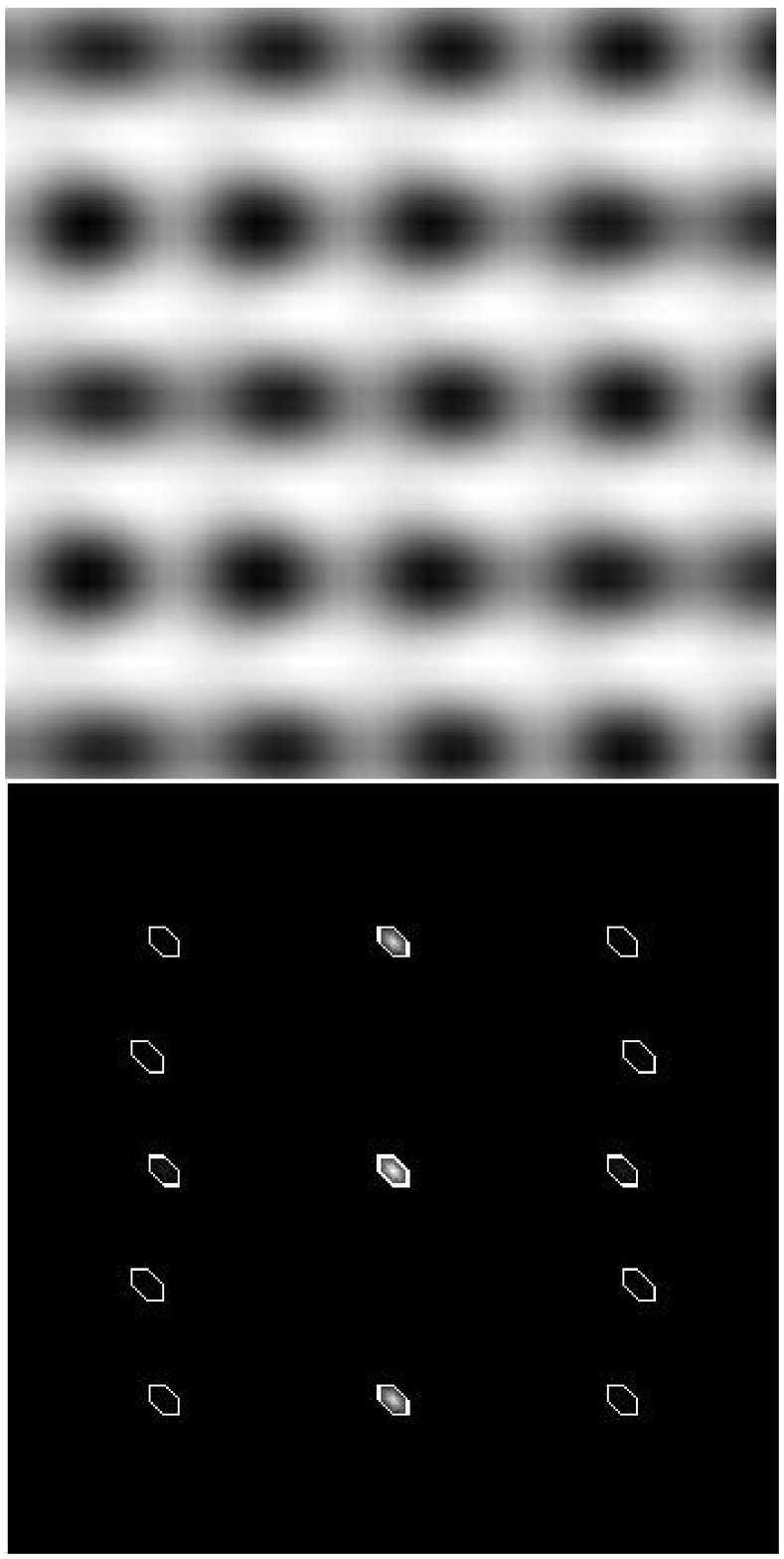}\label{subfig:vo0.0900f0.3860000000}}
\end{center}
\caption{Change of the structure factor with the applied force for $\delta_{\rm{m}}=0.125,V_0=0.0900$, where depinning is continuous. The images correspond to (a) $f=0.3776$ (before the depinning transition marked
by the vertical arrow in Fig.  \ref{subfig:128-28-0.0900v1}), (b) $f=0.3782$ (after the depinning transition). }\label{fig:vo0.0900f}
\end{figure}

Both depinning mechanisms are accompanied by structural changes.
The system changes from a commensurate (1x1)  phase (below critical threshold) to a
distorted hexagonal phase (Figs. \ref{fig:vo0.0350f},\ref{fig:vo0.0900f}).

\subsubsection{Nonlinear Response of the $c(2\times2)$ phase}

\begin{figure}[!h]
\begin{center}
\subfigure[]{\includegraphics[height=55mm,clip=true,angle=0]{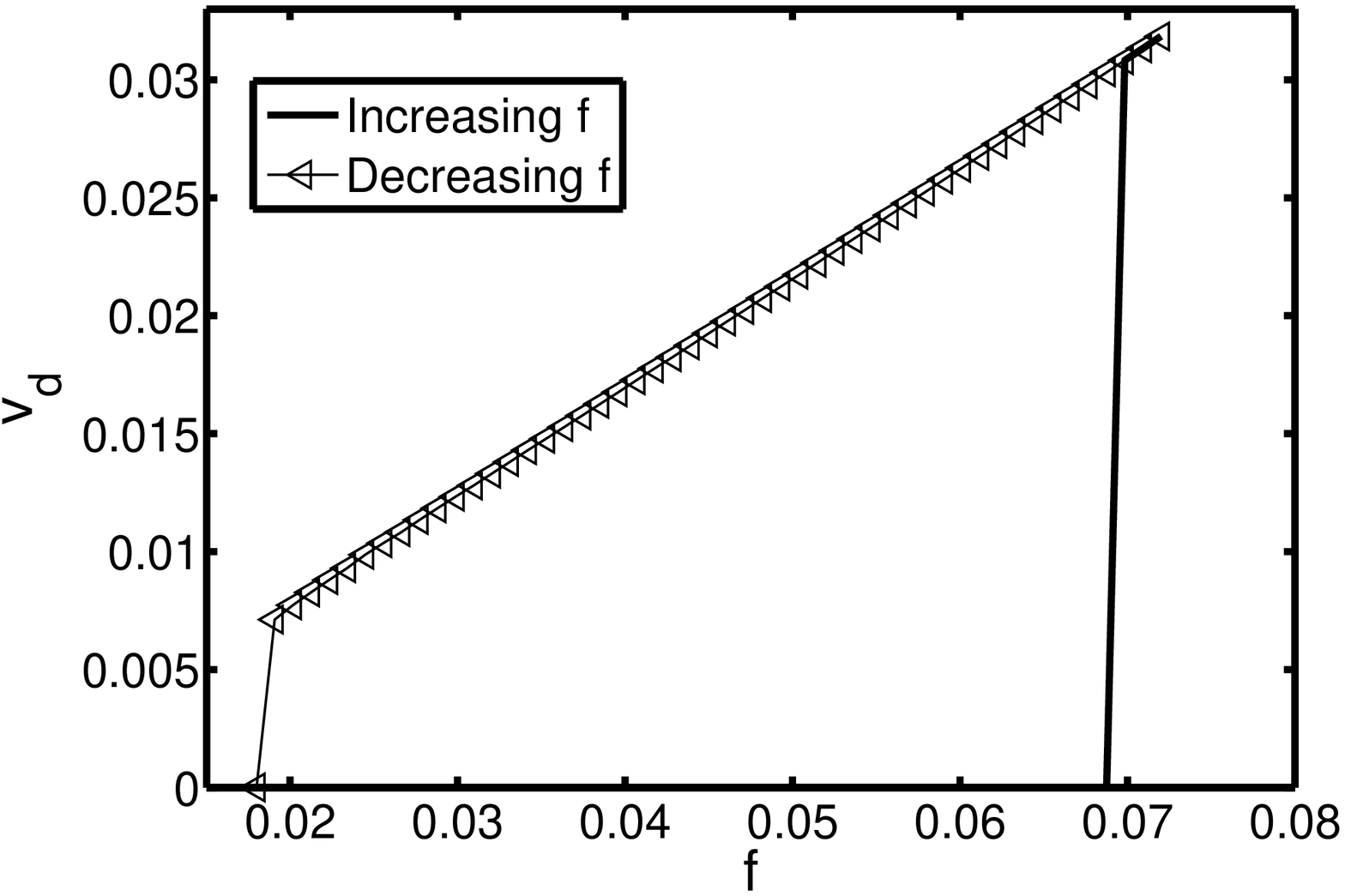}\label{subfig:128-48-0.099v1}}
\subfigure[]{\includegraphics[width=55mm,clip=true,angle=270]{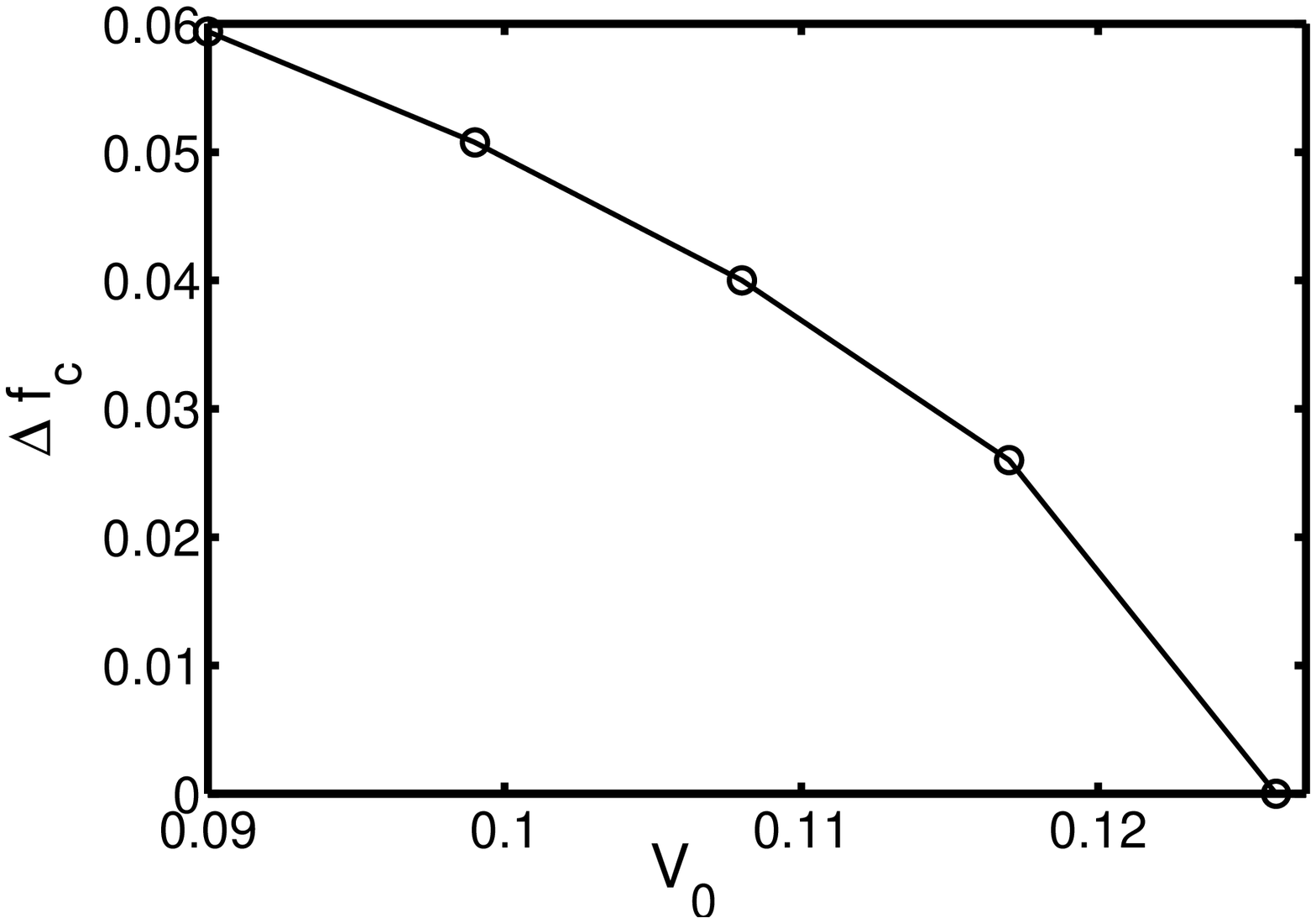}\label{subfig:128-deltafc-28}}
\end{center}
\caption{(a) The variation of the velocity with respect to the
external force for a discontinuous transition for the
$\textrm{c}(2\times 2)$ phase ($\delta_{\rm{m}}=-0.50,V_0=0.099$)
and (b) $\Delta f_c$ vs $V_0$ for $\delta_{\textrm{m}}=-0.50$.
}\label{fig:128-48-0.099}
\end{figure}

\begin{figure}[!h]
\begin{center}
\subfigure[]{\includegraphics[width=55mm,clip=true,angle=270]{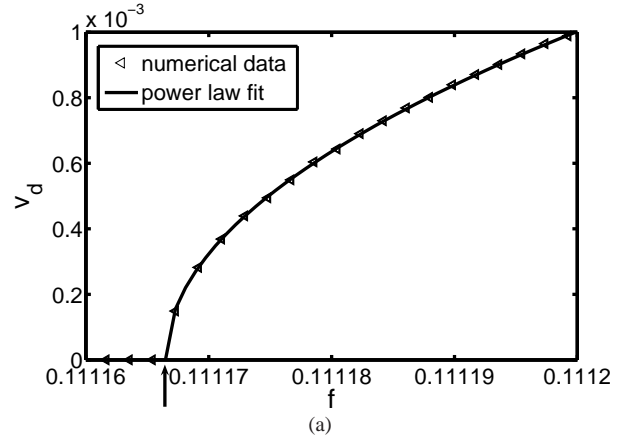}\label{subfig:128-48-0.144v1}}
\end{center}
\caption{\label{fig:128-48-0.144} The variation of the velocity with
respect to the external force for a continuous transition
($\delta_{\rm{m}}=-0.50,V_0=0.207$) for the $\textrm{c}(2\times 2)$ and the corresponding power law fit. The vertical arrow marks the value of the critical depinning force.The triangles represent the numerical data,
while the continuous line the power law fit with $\zeta=0.50\pm 0.03$.}
\end{figure}

\begin{figure}[!h]
\begin{center}
\subfigure[]{\includegraphics[width=35mm,clip=true,angle=0]{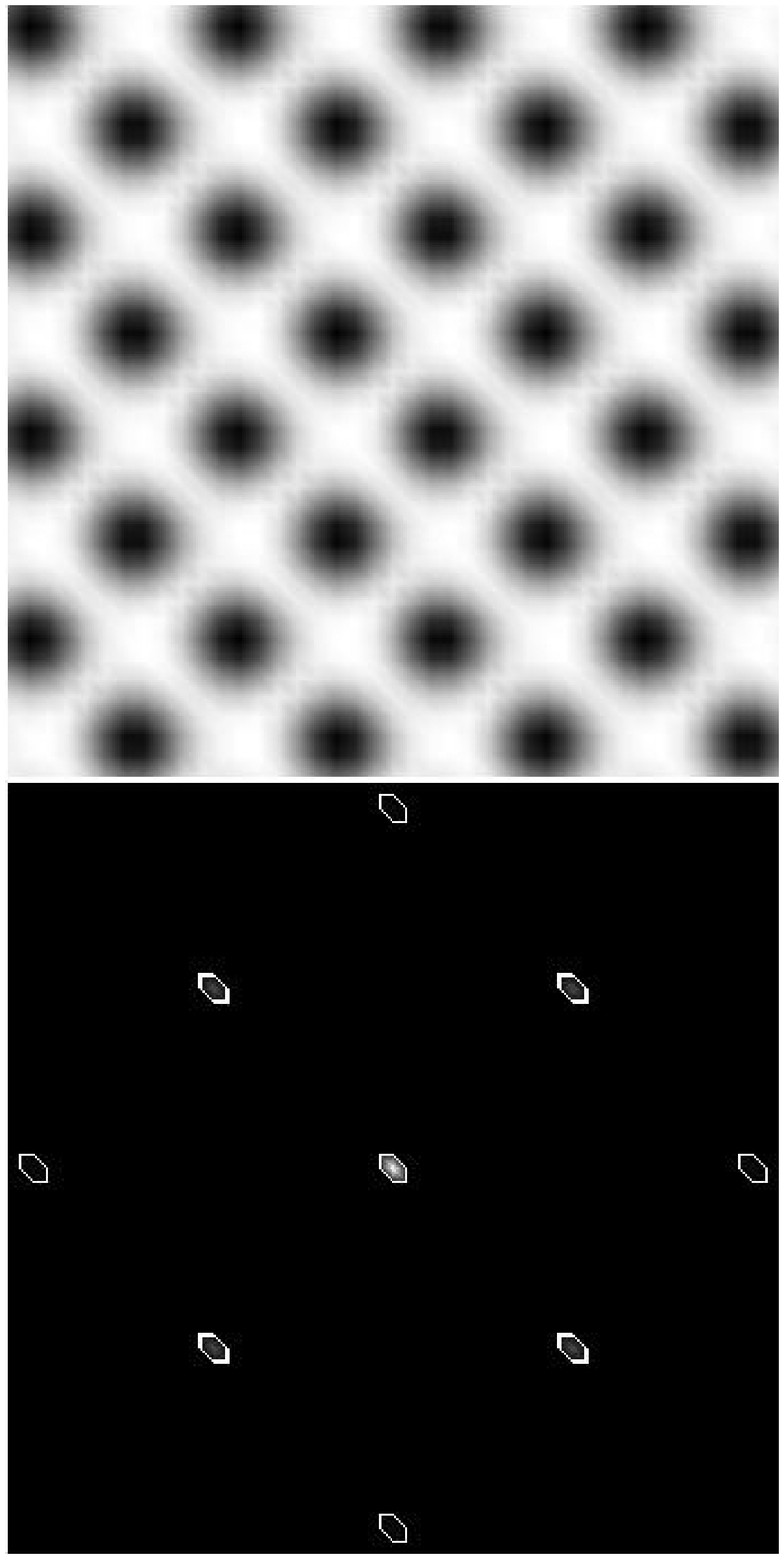}\label{subfig:vo0.099f0.018000000}}
\subfigure[]{\includegraphics[width=35mm,clip=true,angle=0]{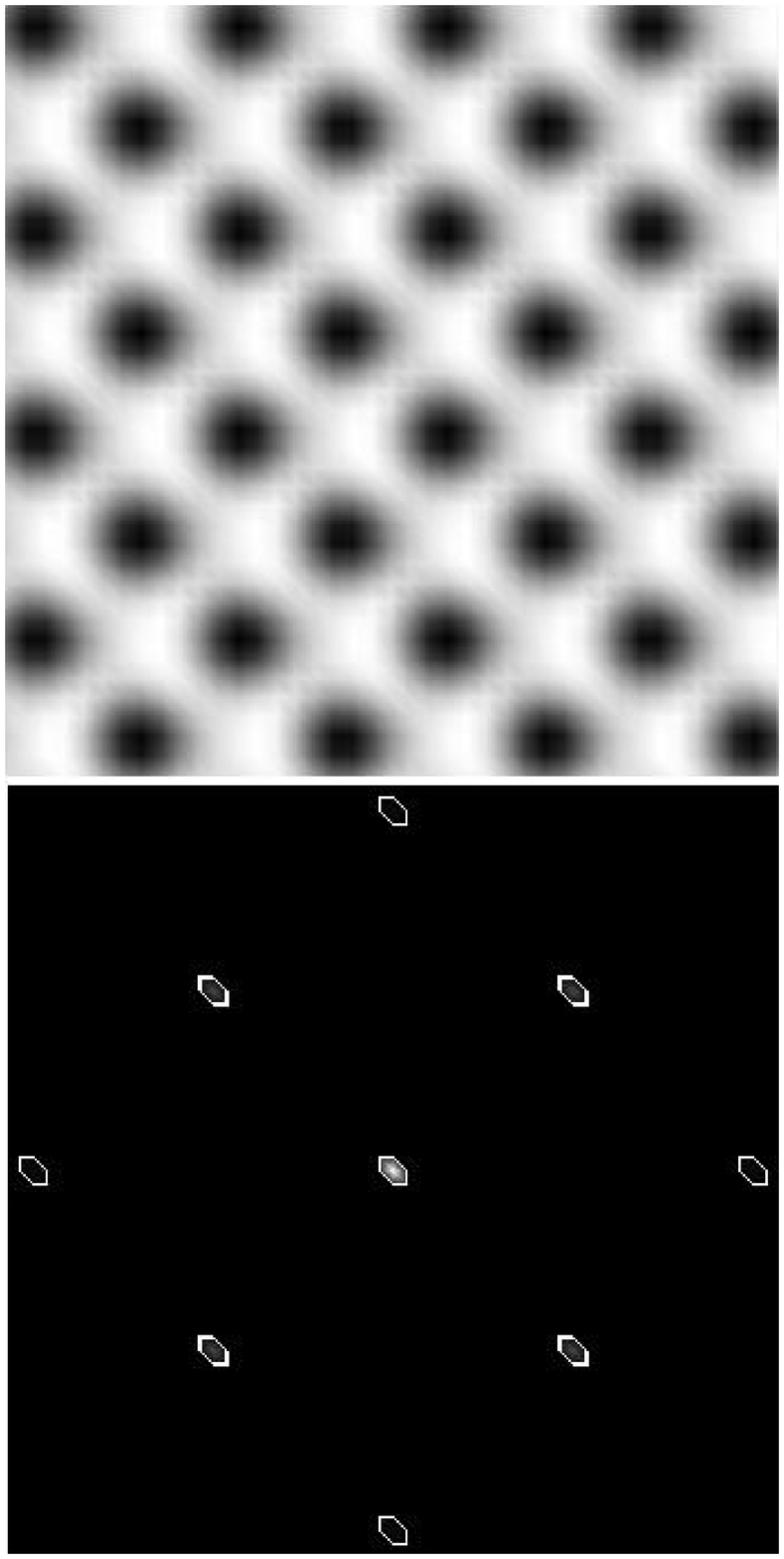}\label{subfig:vo0.0350f0.11840up}}
\subfigure[]{\includegraphics[width=35mm,clip=true,angle=0]{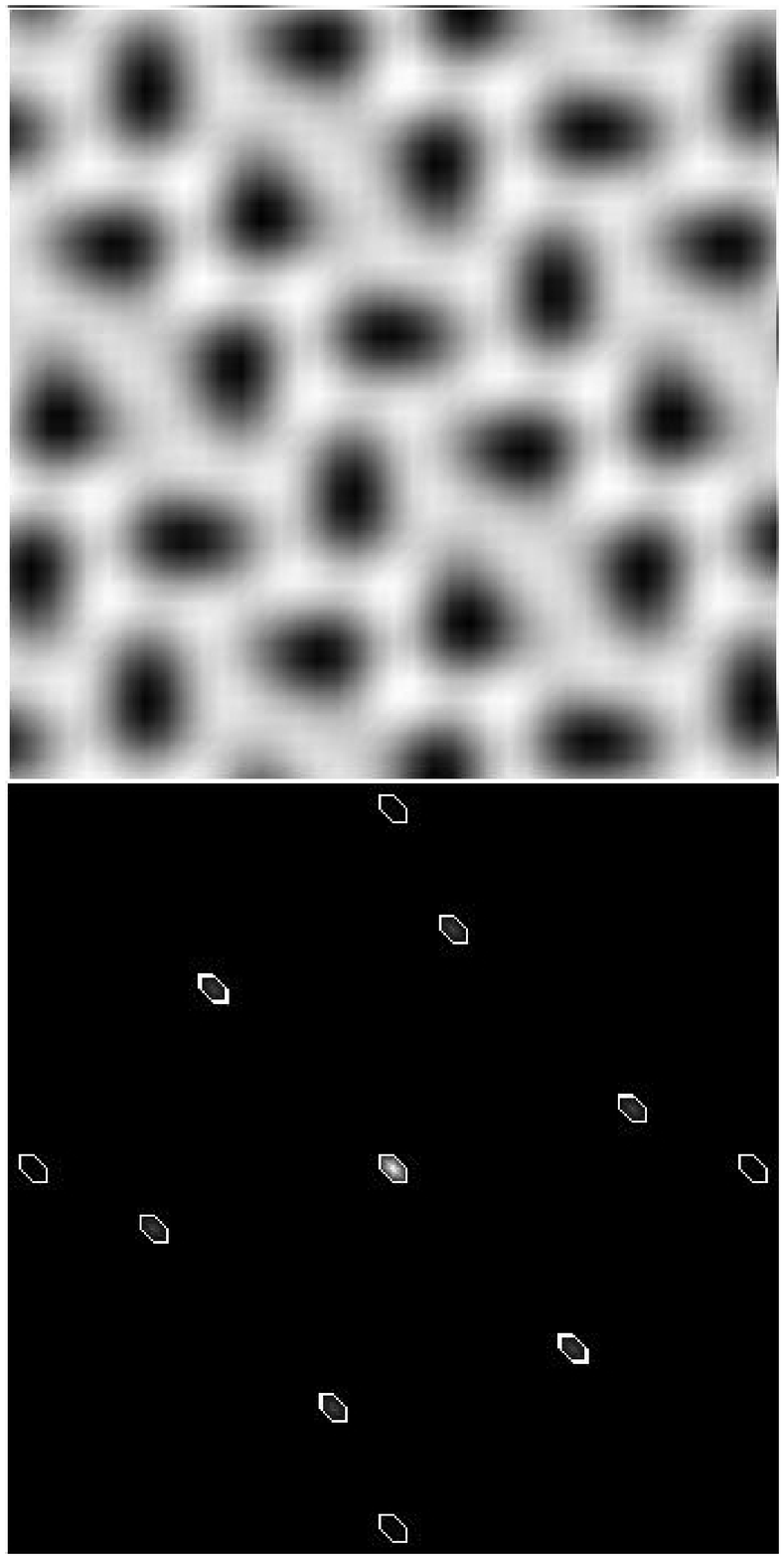}\label{subfig:vo0.0350f0.13040}}
\subfigure[]{\includegraphics[width=35mm,clip=true,angle=0]{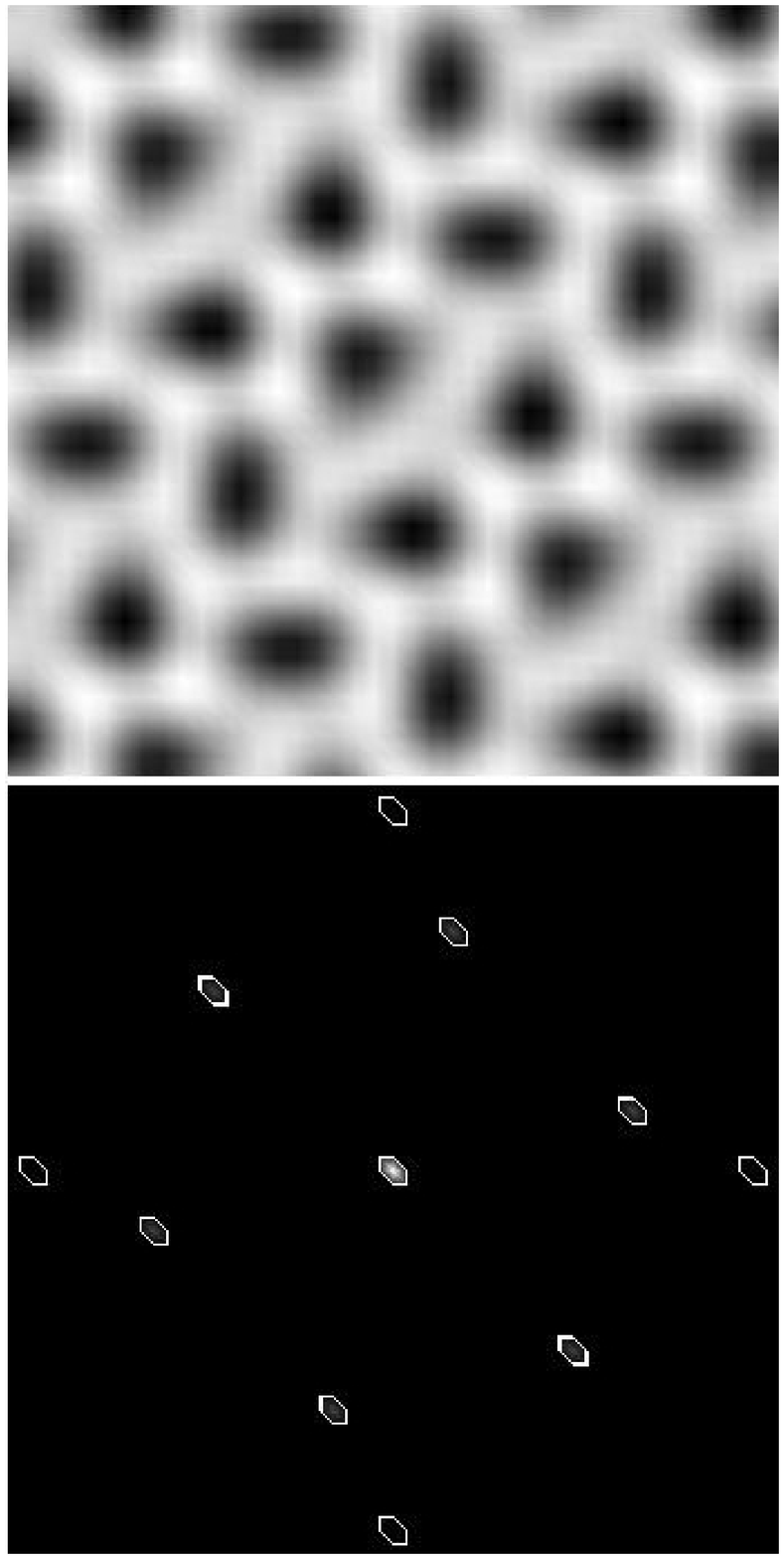}\label{subfig:vo0.0350f0.11000do}}
\end{center}
\caption{Change of the lattice structure (upper panels) and the corresponding structure factor (lower panels) with the applied force for $\delta_{\rm{m}}=-0.5,V_0=0.099$ for the $\textrm{c}(2\times 2)$ ,where depinning
is discontinuous. Image (b) corresponds to $f=0.10504$ with a non-moving initial configuration, while for (c) the applied force is the same but the initial configuration is a moving one. The cases (a) $f=0.018$, (d)
$f=0.13$  are outside of the hysteresis region and same result is obtained with moving or non-moving initial configuration.}\label{fig:vo0.099f}
\end{figure}

\begin{figure}[!h]
\begin{center}
\subfigure[]{\includegraphics[width=35mm,clip=true,angle=0]{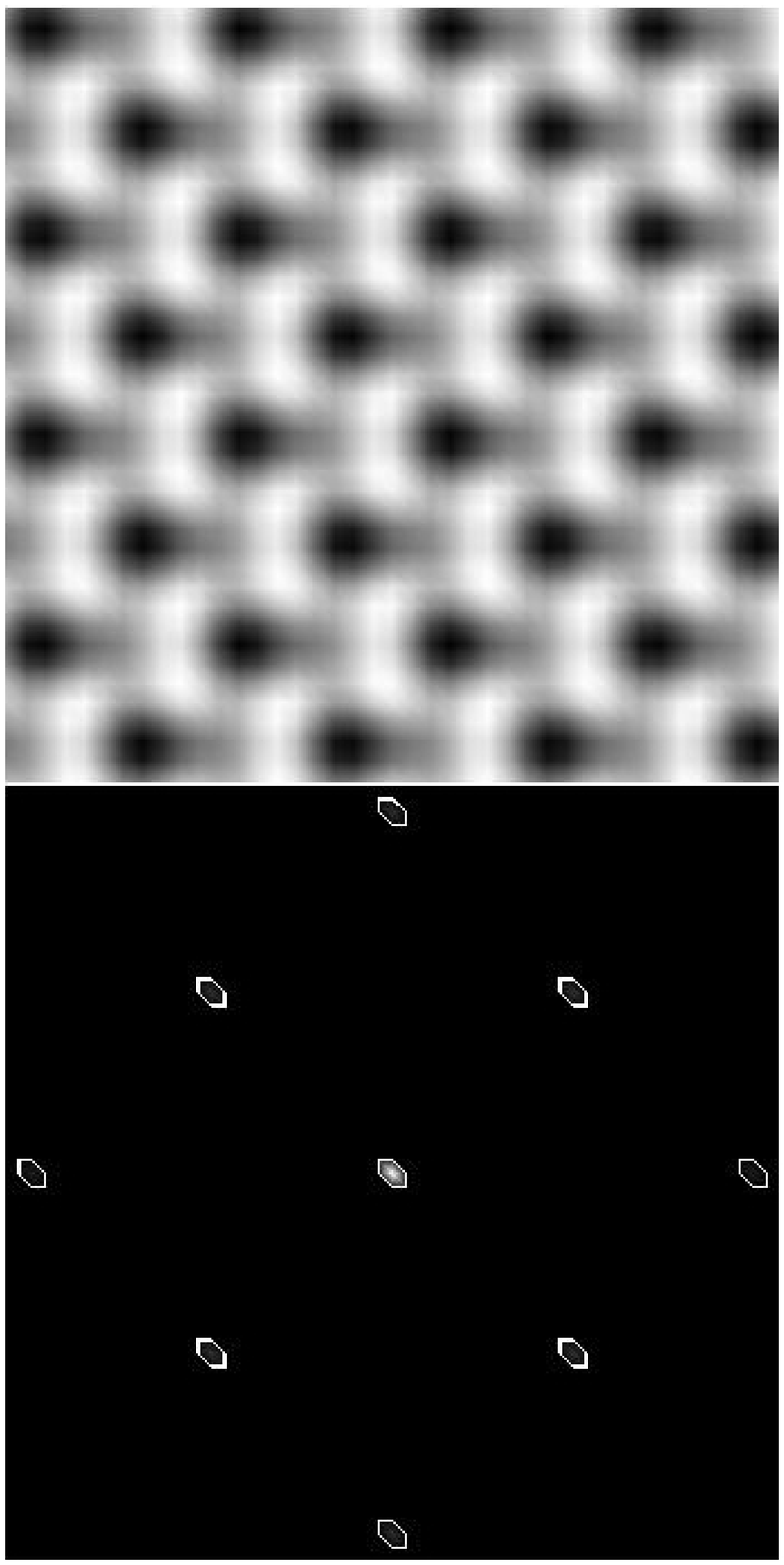}\label{subfig:vo0.207f0.1111664528}}
\subfigure[]{\includegraphics[width=35mm,clip=true,angle=0]{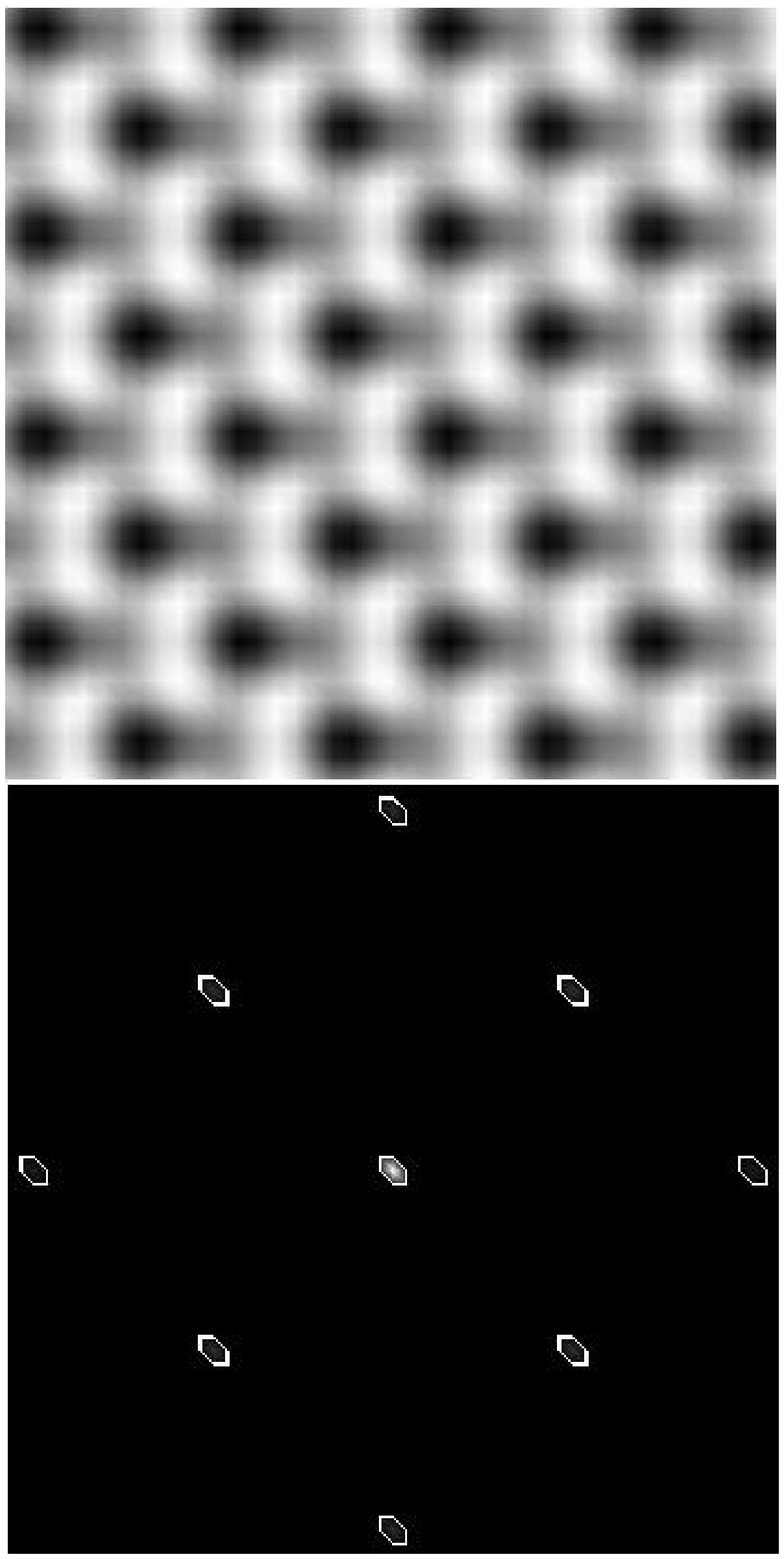}\label{subfig:vo0.207f0.1111664528}}
\end{center}
\caption{Change of the structure factor with the applied force for
$\delta_{\rm{m}}=-0.50,V_0=0.207$ for the $\textrm{c}(2\times 2)$, where depinning is continuous.
The images correspond to a) $f=0.1111664528$ (right before the depinning transition indicated by the vertical arrow in Fig.  \ref{subfig:128-48-0.144v1}, b) $f=0.1111672800$ (right after the depinning
transition).}\label{fig:vo207f}
\end{figure}

Similar to the $(1\times1)$ phase, the commensurate $c(2\times2)$
phase also exhibits both discontinuous and continuous depinning.
For low value of pinning strength discontinuous depinning
transition was found (Fig. \ref{fig:128-48-0.099}), while for
large values of the pinning strength the depinning becomes
continuous (Fig. \ref{fig:128-48-0.144}) with the exponent
$\zeta=0.5$. In both cases the structure of the systems changes
when the force is applied  from a C phase to a distorted hexagonal
depinned phase (Figs. \ref{fig:vo0.099f},\ref{fig:vo207f}).

While for the cases presented here the structural changes occur when the system starts to depin,
it is also possible for changes to occur for forces below the critical threshold.
For some values of the mismatch and pinning strength, if the force is rotated $45$
degrees the $c(2\times 2)$ phase will change first to a $(1\times1)$ phase from which
the systems depins continuously as described before. Other force induced transitions
between commensurate phases with no sliding are also possible.

The hysteresis behavior of the  depinning transition  found for
sufficiently small pinning strength $V_0$ and the critical exponent
$\xi$ for the continuous transition for larger $V_)$ are confirmed
by calculations of peak velocity $v_p$ from Eq. (\ref{velpeak}). For
large $V_0$ where there is no hysteresis, the behavior of $v_p$ as
function of $f$ show a depinnning transition at a critical force
$f_c$. A power-law fit of the velocity near  $f_c$ gives an exponent
$\zeta = 0.52\pm 0.03$ which is consistent with the estimate using
the velocity definition in Eq. (\ref{velgrad}).

\begin{figure}
\includegraphics[ width=9.5 cm]{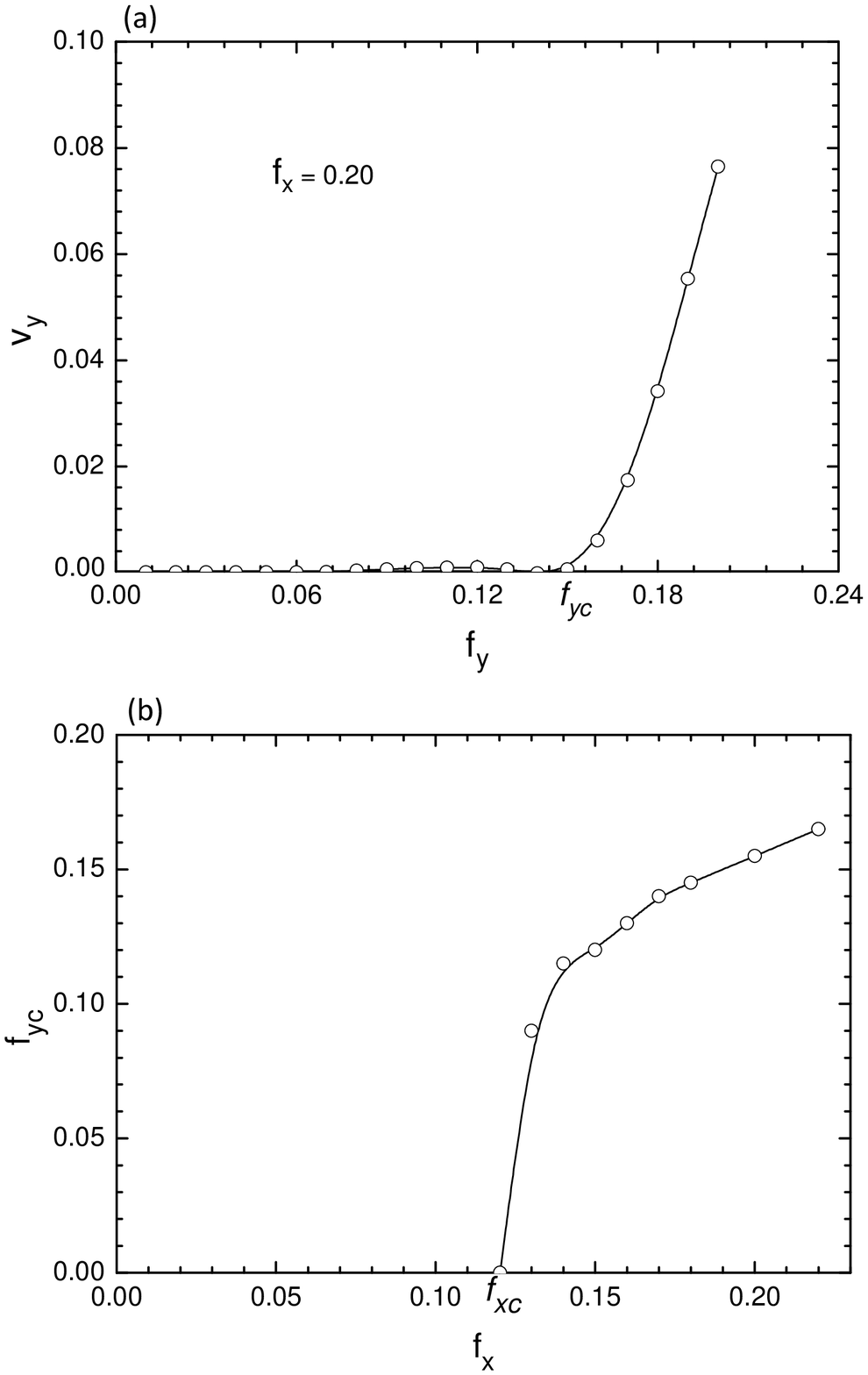}
\caption{(a) $v_y$ as a function of an additional force $f_y$ in
the transverse direction, with $f_x$ fixed. (b) Critical
transverse force $f_{cy}$ as a function of the longitudinal force
$f_x$. Results for $V_0=0.275$, $\delta_m =-0.5$.} \label{trans}
\end{figure}

The determination of the velocity response from the peak positions
allows us also to study the response to an additional force $f_y$
applied perpendicular to the longitudinal force $f_x$ in the moving
state. For $f_x >> f_c$ the longitudinal velocity $v_x$ is
proportional to the force since in the moving state the external
pinning potential in the direction of the force appears as a time
dependent perturbation in a reference system co-moving with the
lattice, with a vanishing time average \cite{bib:Doussal}. However,
the pinning potential remains static in the transverse direction
\cite{bib:Doussal}. One then expects that for sufficiently larger
pinning strength, a transverse depinning transition is possible for
increasing force $f_y$ while $f_x$ is kept fixed. Fig.
\ref{trans}(a) shows the behavior of the transverse velocity
component $v_y$ when an increasing $f_y$ is applied in the moving
state with fixed $f_x > f_c$. The transverse critical force $f_{yc}$
decreases with the longitudinal force $f_x$ and appears to vanishes
at the longitudinal depinning transition $f_c$, as shown in Fig.
\ref{trans}(b).

\section{Discussion and Conclusions}

In this work we have considered the recently developed phase field
crystal model \cite{bib:Elder04rq} in the presence of an external
periodic pinning potential \cite{bib:Achim06au,bib:Achim08ap} and a
driving force. As the model naturally incorporates both elastic and
plastic deformations, it provides a continuum description of lattice
systems such as adsorbed atomic layers on surfaces or 2D vortex
lattices in superconducting thin films, while still retaining the
discrete lattice symmetry of the solid phase. The main advantage of
the model as compared to traditional approaches is that despite
retaining spatial resolution at its lowest length scale its temporal
evolution naturally follows diffusive time scales. Thus the
numerical simulation studies of the dynamics of the systems can be
achieved over realistic time scales, which for example in the case
of adsorbed atomic systems  may correspond to many orders of
magnitude over the time scale used in microscopic atomic models. In
this work we have exploited this method to determine the phase
diagram in one and two dimensions as a function of lattice mismatch,
pinning strength and a driving force.

We have concentrated on the nonlinear response to an external driving force
on the most common stable commensurate states, namely the $(1\times1)$
and the $c(2\times2)$ phases. These are particularly interesting cases which
are relevant for physical systems of current interest and accessible
experimentally such as, driven adsorbed layers
\cite{bib:Perssonbook,bib:Persson93}, which determine the sliding friction
behavior between two surfaces with a lubricant \cite{bib:Israelechvili} and
between adsorbed layers and an oscillating substrate
\cite{bib:Krim,bib:Mistura}.

Our results for the phase field crystal model with overdamped
dynamics indicate both discontinuous and continuous transitions
depending on the magnitude of the pinning strength. For high enough
pinning strengths continuous transitions occurred with the velocity
near the transition scaling as $(f-f_c)^{1/2}$, independent of the
dimension of the system. This is as expected, since for a
commensurate state in  a strong periodic pinning potential, each
'particle' acts independently and the model reduces to an effective
single particle in a periodic potential, with a known depinning
exponent of $1/2$. Perhaps more interesting is the observation of
discrete transitions and hysteresis loops found at low pinning
strengths. In the two-dimensional case, the observed hysteresis
behavior is consistent with the arguments and atomistic molecular
dynamics simulations of driven adsorbed layers
\cite{bib:Perssonbook,bib:Persson93}  indicating that hysteresis
remains in the overdamped limit. However, our results show that it
disappears for large enough pinning strength.  For the discontinuous
transition, there are two different critical values, $f_{c}^{in}>
f_{c}^{de}$, correspond to the static and kinetic critical forces,
respectively, which lead the to stick and slip motion at low sliding
velocities as observed experimentally \cite{bib:Israelechvili}. The
general features observed of hysteresis and power-law behavior near
the continuous depinning transition are also of interest for driven
charge density waves \cite{bib:Gruner81,bib:Kart99} and driven flux
lattices \cite{bib:Nori99,bib:Blatter94} although in these cases
there are important additional effects due to a large degree of
disorder in the pinning potential. Whether the present phase field
model also shows the same main features observed for the atomistic
model in presence of thermal fluctuations
\cite{bib:Persson93,bib:Perssonchem,bib:Granato00} is an interesting
question which will require further investigation.

In addition to the longitudinal depinning transition where the
lattice system is moving in the same direction as the driving force,
a driven two-dimensional lattice on a periodic potential can also
show an interesting behavior for the  transverse response in the
moving state. When the lattice is already moving along some symmetry
direction of the pinning potential the response to an additional
force applied in the direction perpendicular to the longitudinal
driving force may lead to a depinning transition for increasing
transverse force \cite{bib:Doussal}. Such transverse depinning has
been found in different driven lattice systems with periodic pinning
including driven vortex lattices \cite{bib:Nori99,bib:reichhardt08}  and adsorbed
layers \cite{bib:Granato00} in standard molecular dynamics
simulations. In the present PFC model, we have obtained similar
results for the transverse depinning. Experimentally, some evidence
of transverse pinning has been observed in measurements on
charge-density waves \cite{bib:cdw}, Wigner solid \cite{bib:wigner}
and vortex lattices \cite{bib:vlatt}, although in these cases
disorder in the pinning potential plays a more important role.

While the hysteresis behavior is expected in the presence of
inertial terms both in 1D and 2D, it is quite interesting to see it
in 1D when the dynamics being used is overdamped and purely
relaxational. In 2D, it can be argued that topological defects such
as dislocations can lead to this behavior even with overdamped
dynamics but these defects are not available in 1D. It is
interesting to speculate that the hysteresis behavior is intimately
related to the need for plastic deformations to mediate the
transition from one lattice structure to another. Work on these
problems is already in progress.

\begin{acknowledgments}
This work was supported by joint funding under EU STRP 016447 MagDot
and NSF DMR Award No. 0502737 (C.V.A. and T.A-N.). Computations were
performed in the  CSC's computing environment. CSC is the Finnish IT
Center for Science and is owned by the Ministry of Education. E.G.
was supported by Funda\c c\~ao de Amparo \`a Pesquisa do Estado de
S\~ao Paulo - FAPESP (Grant no. 07/08492-9).  K.R.E. acknowledges the support
from NSF under Grant No.  DMR-0413062. MK has been supported by
the Natural Sciences and Engineering Research Council of Canada (NSERC)
and SharcNet (www.sharcnet.ca).
\end{acknowledgments}

\bibliographystyle{apsrev}

\end{document}